\documentclass[a4paper,10pt]{article}
\usepackage{graphicx}
\usepackage{epsfig,a4}

\title{Diffractive scattering at high energies \footnote{This report was pesented on the jubilee scientific session-conference of Nuclear physics section Russioan Academy of sciences (5-9th of December, 2005)}}
\author{Abramovsky V.A. \footnote{ava@novsu.ac.ru}, Dmitriev A.V. \footnote{gridlab@novsu.ac.ru}, Schneider A.A. \\ \footnotesize{ Novgorod State University, B. S.-Peterburgskaya Street 41,}\\ \footnotesize{Novgorod the Great, Russia, 173003} }
\begin{document}

\maketitle

\begin{abstract}
In this work we analyse applicability of regge-based models to hadron interactions. Commonly-used extensions of pure regge model are found to be failed. Low constituent model is inroduced and found to be consistent with hadron-hadron and hadron-photon data.
\end{abstract}

\section{Introduction}

In this work we review contemporary methods for description of diffraction processes at high energy. In the soft region of kinematical variables QCD has not been solved, so, phenomenological approaches are traditionally used. 

The key object of investigation in such approaches is Pomeron. Pomeron is defined as reggeon with largest intercept at canonical regge model or as synonim of diffraction ("diffracton") in more contemporary context. First context of pomeron can be used as basis for the second context, and we get "full" eikonalized pomeron made as the sum of n-pomeron cuts. Another way is to write up propagator of pomeron and vertex pomeron-hadron and pomeron-pomeron functions, fitted to experimental values.

In this work we consider existing models of pomeron and compare them with our low contituent model.

In section \ref{sec:quasi} we investigate applicability of the quasi-elastic model for the description of hadron-hadron and hadron-photon diffractiventeractions. 

Main goal of subsection \ref{sub:totel} section is to estimate eikonal parameter $c$ from the elastic and total cross-section data.

In subsection \ref{sub:quasi}  we consider processes of the single diffraction in the framework of the quasi-eikonal model. As compared  with Ref.\cite{gotsman_orig} and Ref.\cite{chung_orig} we consider all non-enhanced diagrams.

In this work we also consider low constituent model (section \ref{sec:lcm}) , where there is only basic quark-gluon states and interactions. This model leads us to the non-local pomeron, but it has clear interpretation of the pomeron flux renormalization.

Applicability of considered models to photon-hadron interections is discussed in section \ref{sec:photon}.

\section{Main problems to be solved}
\label{sec:main}

\begin{figure}
\includegraphics[scale=0.35]{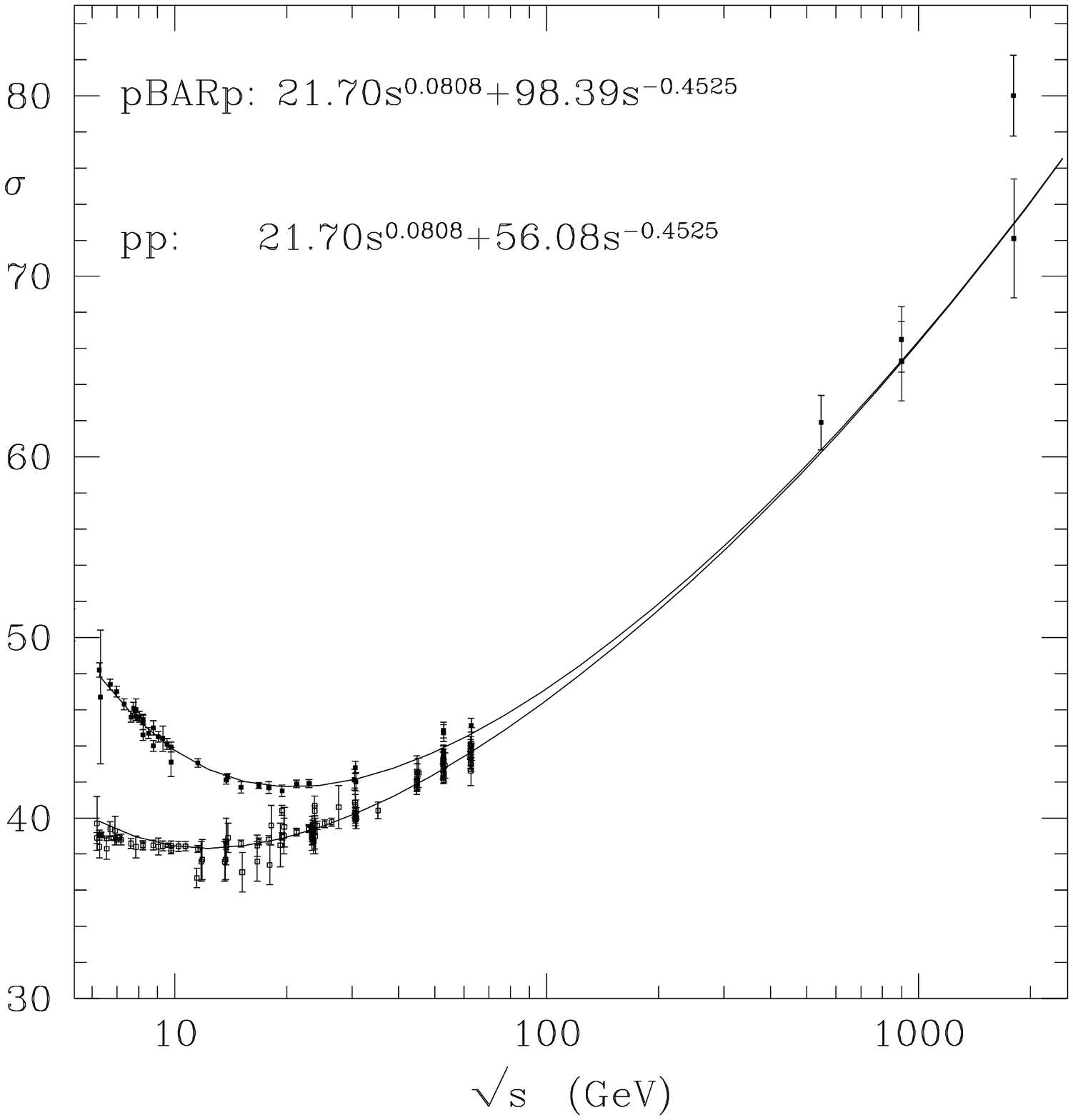}
\includegraphics[scale=0.35]{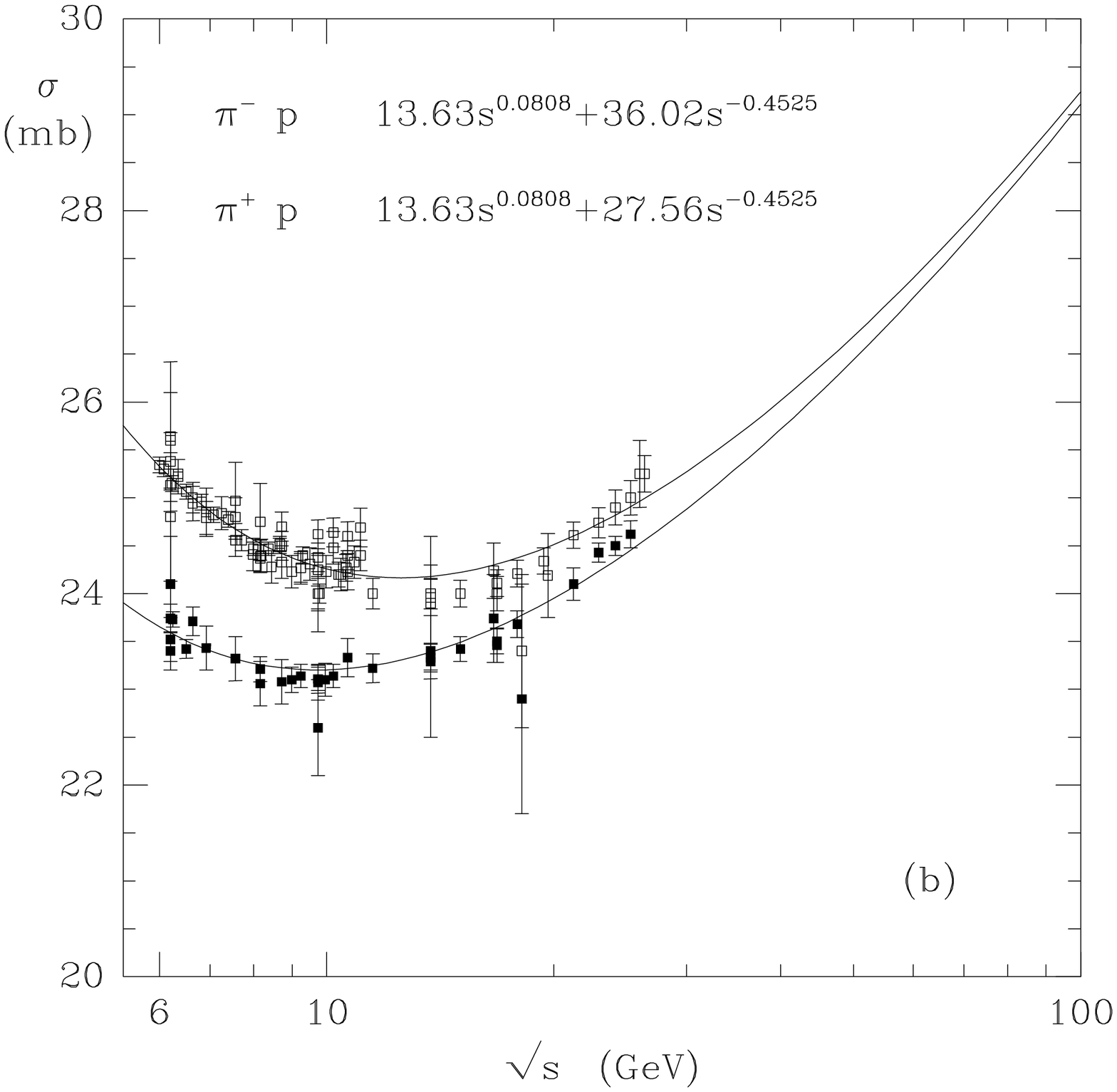}
\label{fig:sigtot}
\caption{Total cross-sections in the Donnachie-Landshoff parameparameterization}
\end{figure}

\begin{figure}
\begin{center}                                                               
\begingroup%
  \makeatletter%
  \newcommand{\GNUPLOTspecial}{%
    \@sanitize\catcode`\%=14\relax\special}%
  \setlength{\unitlength}{0.1bp}%
\begin{picture}(4679,2807)(0,0)%
\special{psfile=fig/dsigdt41_alls.pst llx=0 lly=0 urx=936 ury=655 rwi=9360}
\put(2293,724){\makebox(0,0)[l]{\small23.5 GeV}}%
\put(2293,1073){\makebox(0,0)[l]{\small30.7 GeV $(\times 100)$}}%
\put(2293,1444){\makebox(0,0)[l]{\small44.7 GeV $(\times 10^4)$}}%
\put(2293,1790){\makebox(0,0)[l]{\small52.8 GeV $(\times 10^6)$}}%
\put(2293,2181){\makebox(0,0)[l]{\small\llap{$\sqrt s=\;$}62.5 GeV $(\times 10^8)$}}%
\put(1839,50){\makebox(0,0){$-t$ [GeV$^2$]}}%
\put(200,1568){%
\special{ps: gsave currentpoint currentpoint translate
270 rotate neg exch neg exch translate}%
\makebox(0,0)[b]{\shortstack{$d\sigma/d t$ [mb/GeV$^2$]}}%
\special{ps: currentpoint grestore moveto}%
}%
\put(3078,229){\makebox(0,0){2.2}}%
\put(2803,229){\makebox(0,0){2}}%
\put(2527,229){\makebox(0,0){1.8}}%
\put(2252,229){\makebox(0,0){1.6}}%
\put(1977,229){\makebox(0,0){1.4}}%
\put(1701,229){\makebox(0,0){1.2}}%
\put(1426,229){\makebox(0,0){1}}%
\put(1151,229){\makebox(0,0){0.8}}%
\put(875,229){\makebox(0,0){0.6}}%
\put(600,229){\makebox(0,0){0.4}}%
\put(550,2807){\makebox(0,0)[r]{$10^8$}}%
\put(550,2453){\makebox(0,0)[r]{$10^6$}}%
\put(550,2099){\makebox(0,0)[r]{$10^4$}}%
\put(550,1745){\makebox(0,0)[r]{100}}%
\put(550,1391){\makebox(0,0)[r]{1}}%
\put(550,1037){\makebox(0,0)[r]{0.01}}%
\put(550,683){\makebox(0,0)[r]{$10^{-4}$}}%
\put(550,329){\makebox(0,0)[r]{$10^{-6}$}}%
\end{picture}%
\endgroup
 
\end{center}
\caption{The Donnachie-Landshoff fit for the differential elastic $pp$ cross section}
\label{fig:sigel}
\end{figure}
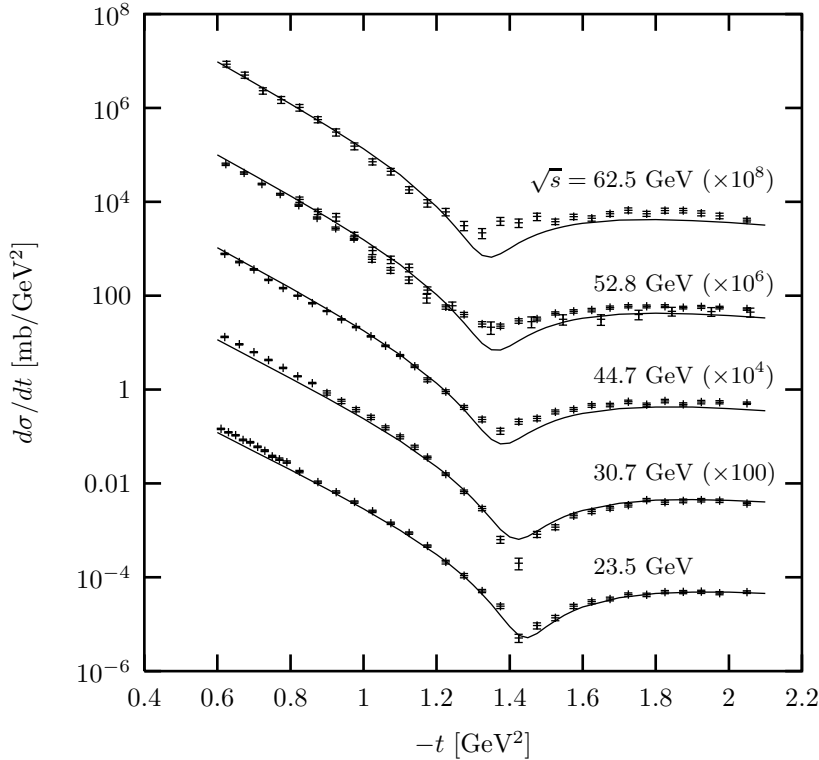

Regge non-enhanced phenomenology well describes total and elastic cross-sections in the Donnachie-Landshoff parameterization{DL}, see Fig.\ref{fig:sigtot}, taken from \cite{DLfig} and Fig.\ref{fig:sigel}, taken from  \cite{elfig}.

This model has problems with initarity.
At first, total cross-sections rises too fast, breaking Fruassar limit \cite{fruassar}:
\begin{equation}
\sigma_{tot}=\frac{1}{s}ImA(s,0) < \frac{\pi}{m_{\pi}^2}ln^2(s)
\label{Fruassar}
\end{equation}
At contemporary collider energies this fact is not very annoying, because observable cross-sections is much lower Fruassar limit.

But initarity is break up at $s$-chanel, where we have limit on profile function $J(s,b)$:

\begin{equation}
J(s,b)= 2Im\ \chi^{\bar pp}_{pp} (s,b) < 1
\end{equation}

\begin{figure}
\begin{center}
\includegraphics[scale=1]{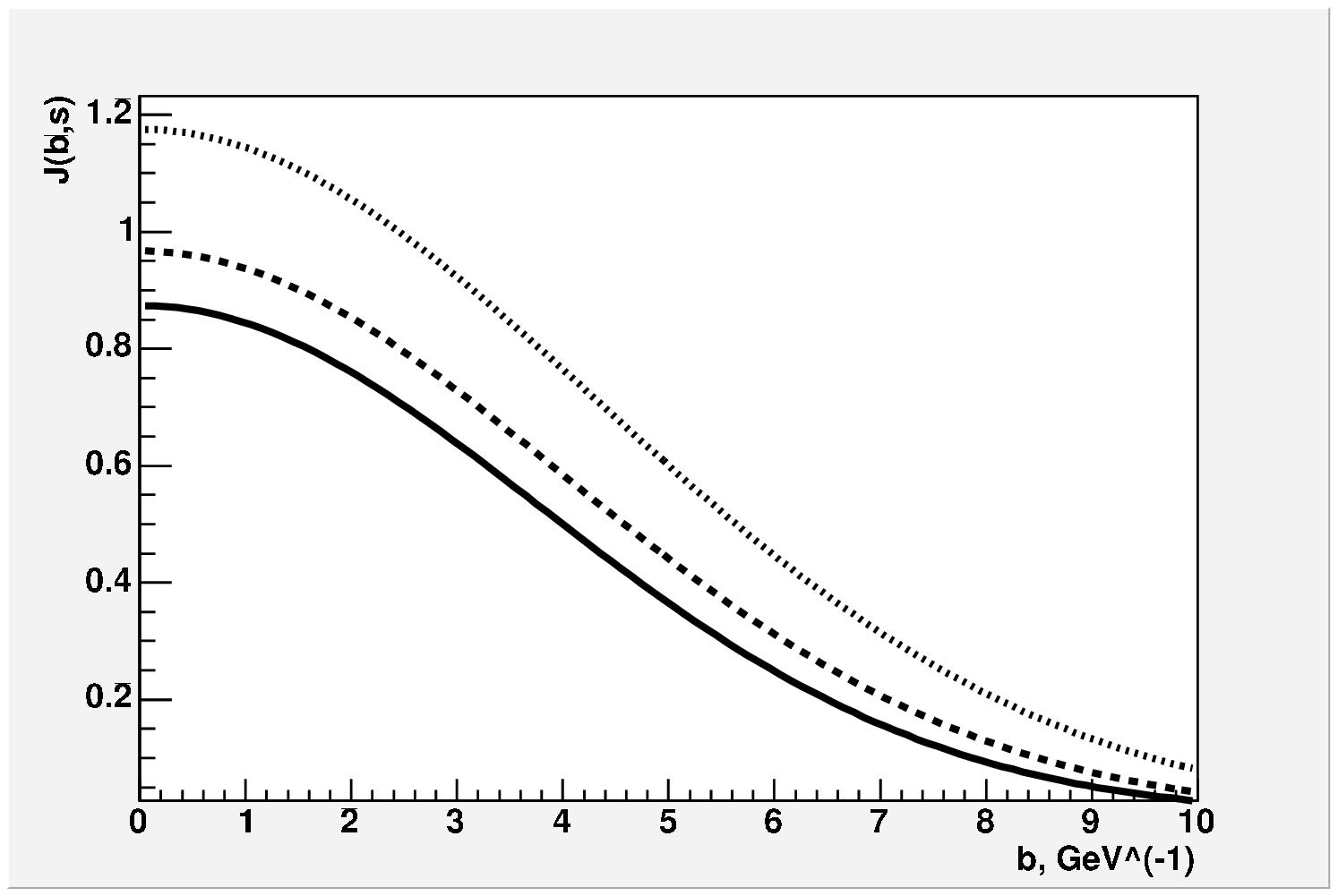}
\caption{Profile function $J(b,s)$ at $\sqrt(s)=546 GeV$ (solid line), $\sqrt(s)=1800 GeV$(dashed line), $\sqrt(s)=14 TeV GeV$ (dotted line).}
\label{fig:profile_Regge}
\end{center}
\end{figure}

Reconstruction of profile function shown at Fig.\ref{fig:profile_Regge}

It is clear, that pure Regge model break up unitarity already at LHC energies.

Unitarity at $s$-chanel is closely connected with sum rule for elastic cross-sections derived in \cite{Pancheri:2004ct}:

\begin{equation}
	\begin{array}{l}
		I_0(s) = {{1}\over{2}}\int_{-\infty}^0 \sqrt{{{d\sigma}\over{\pi dt}}} dt
		\ \rightarrow\ 1
		\\
		s \rightarrow \infty
	\end{array}
	\label{eq:i0s}
\end{equation}

\begin{figure}
	\begin{center}
		\includegraphics[scale=0.5]{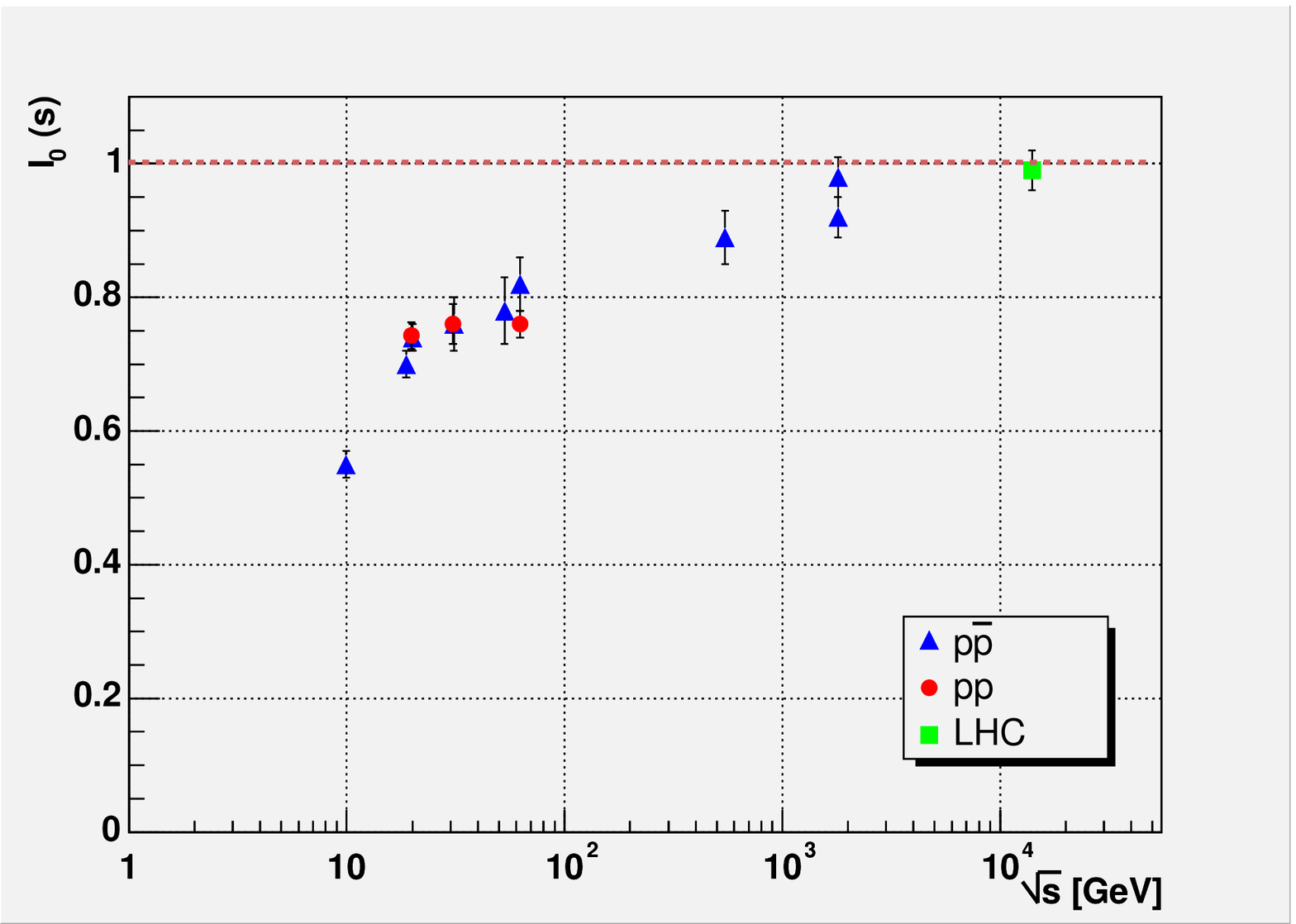}
	\end{center}
	\vspace*{0pt}
	\caption{Dependence of $I_0(s)$ (eq.(\ref{eq:i0s})) on $\sqrt s$, last point - extrapolation for LHC. Figure is gotten from \cite{Pancheri:2004ct}.}
	\label{fig:asimpt_I0}
\end{figure}

One can see from Fig.\ref{fig:asimpt_I0}, $I_0(s)$ close to unity at Tevatron energies.

\begin{figure}
\begin{center}
\includegraphics[scale=0.5]{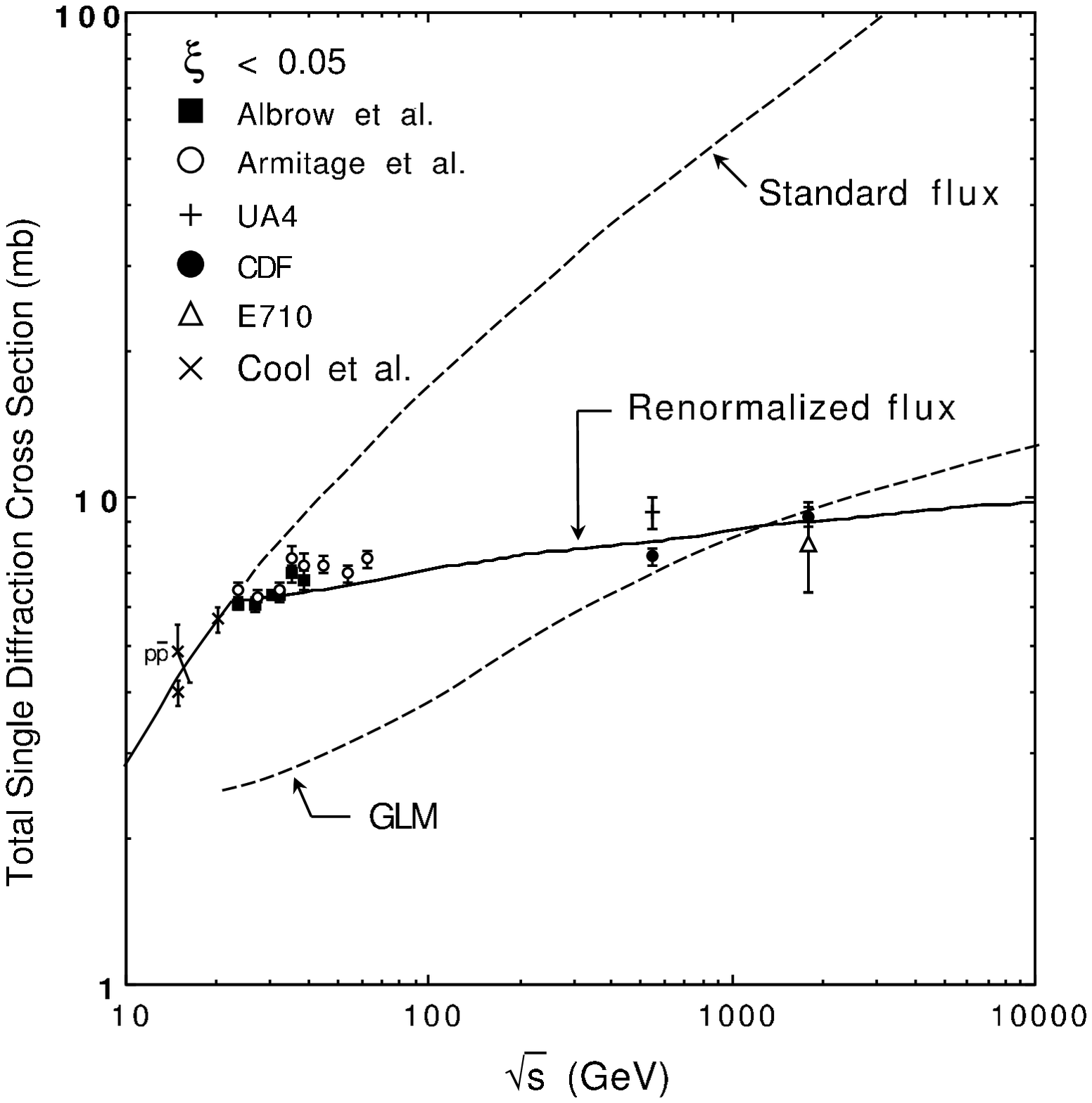}
\vspace*{0pt}
\caption{The total single diffraction cross section for $p(\bar p)+p\rightarrow p(\bar p)+X$ vs $\sqrt s$ compared with the predictions of the renormalized pomeron flux model of Goulianos~\cite{goul_orig} (solid line) and the model of Gotsman, Levin and Maor~\cite{gotsman_orig} (dashed line, labeled GLM); the latter, which includes "screening corrections", is normalized to the average value of the two CDF measurements at $\sqrt s=546$ and 1800 GeV.}
\label{fig:figgoul}
\end{center}
\end{figure}

Low-energy single diffraction data is also well described by regge phenomenology with supercritical pomeron, but at the region of Tevatron energies it fails to describe data on single diffraction dissociation. The main problem is that total single diffraction cross-section rise considerably weaker than it is predicted by $Y$-like Regge diagrams involving only three pomerons. This fact is clearly seen from Fig.\ref{fig:figgoul} extracted from Ref.\cite{goul_pic}, there "Standard flux" corresponds to $Y$-like Regge diagram.

Many ways were suggested to solve this problem. First way is two-variant (Ref.\cite{goul_orig} and Ref.\cite{erhan_orig}) pomeron flux renormalization model, where we consider the equation for cross section of single diffraction
\begin{equation}
\frac{d^3\sigma}{dM^2dt}=f_{I\!\!P /p}(x,t)\sigma_{I\!\!Pp}(s)
\label{triple_gen}
\end{equation}
and pick out the factor, named as 'pomeron flux'
\begin{equation}
f_{I\!\!P /p}(x,t)=K\xi^{1-2\alpha_{I\!\!P}(t))}
\end{equation}
Renormalization of the pomeron flux is made by inserting dependence of $K$ either on $s$ (as in Ref.\cite{goul_orig}) either on $x$,$t$, as  in Ref.\cite{erhan_orig}.
This phenomenological approach well describes CDF data on single diffraction, but we need more theoretical bases for extrapolation to higher energies.

The second way is straight-forward account of screening corrections (Ref.\cite{gotsman_orig} and Ref.\cite{chung_orig}).
This way seems to be more natural, but we need to introduce additional parameters and make some assumptions about Regge
diagram technics. In Ref.\cite{gotsman_orig} and Ref.\cite{chung_orig} only some parts of sufficient diagrams  were accounted by going to the impact parameter space $b$ and replacement initial "Borhn" factor $\chi(s,\overline{b})$ to eikonalized amplitude $(1-e^{-\mu\chi(s,\overline{b})})$. In addiction, the central $Y$-like diagram was modified to account low-energy processes and in Ref.\cite{chung_orig} the dependence of pomeron intercept on energy was introduced.

\section{Applicability of the quasi-eikonal model for the description of hadron-hadron diffractive interactions}
\label{sec:quasi}
\subsection{Estimation of eikonal parameter $c$ from the elastic and total cross-section data.}
\label{sub:totel}

Let`s consider processes of elastic scattering in the framework of the eikonalization procedure. 

There is two view on the eikonalization procedure. At first, eikonalization was considered as a natural way to come from Regge behavior $\sigma_{tot}(s) \sim s^{\Delta}$ to Froussoir-like behavior $\sigma_{tot}(s) \sim ln(s)^2$. Both pure Regge model and black-disk model well describe total cross-section data. From the other hand, data on elastic cross-sections $\frac{d\sigma}{dt}(s,t)$ can be described assuming only one Regge pole (pomeron) at low and moderate $|t|<1GeV^2$. Here we will investigate possibility of eikonalization model to describe elastic scattering data at non-zero transferred momenta.

It is convenient to write up eikonalized amplitude as
\begin{equation}
\begin{array}{l}
A(s,\overline{k})=\frac{s}{i}\frac{4\pi}{c^2}\int \left\{e^{ic\chi(b)}-1\right\}J_0(bk)bdb\\
\chi(b)=\frac{1}{s}\int a(s,\overline{k})J_0(bk)bdb
\end{array}
\end{equation}
Here $\chi(b)$ is eikonal profile function and $a(s,\overline{k})$ is generic amplitude.

In the Regge model eikonal profile $\chi(b)$ can be calculated as
\begin{equation}
\chi(b)=-\frac{c^2}{8\pi}\frac{g_ag_b}{R_a^2+R_b^2+\alpha^{\prime}Y+i\pi\alpha^{\prime}/2}e^{-\frac{1}{4}\frac{b^2}{R_a^2+R_b^2+\alpha^{\prime}Y+i\pi\alpha^{\prime}/2}}e^{\Delta(Y+\frac{i\pi}{2})}
\label{chiR}
\end{equation}
To estimate eikonalization parameters, we will try to describe experimental data on total and elastic cross sections
\begin{equation}
\begin{array}{l}
\sigma_{tot}=\frac{1}{s}ImA(s,0)\\
\frac{d\sigma}{dt}=\frac{1}{16\pi s^2}|A(s,\overline{k})|^2\\
t=\overline{k}^2\\
\end{array}
\end{equation}
at the domain of $0.2GeV^2<t<1GeV^2$ and $\sqrt{s} \geq 53GeV^2$, there we have no non-vacuum reggeons and no perturbative $\frac{1}{t^{10}}$ bebehavior

At first stage we will start from conventional Regge picture, and will vary all parameters in (\ref{chiR}) at given $c$. We take only $\frac{d\sigma}{dt}$ data here to concentrate on the $t$ dependence of $A(s,t)$ and to avoid problems with phase of $A(s,t)$. As the result, we get function of likely-hood $\frac{\chi^2}{n.d.f.}$,  depending on $c$, see Fig.\ref{fig:chi_c}.
\begin{figure}
\begin{center}
\includegraphics[scale=0.75]{fig/chi.epsi}
\end{center}
\caption{Dependence of likelyhood function on c}
\label{fig:chi_c}
\end{figure}

\begin{figure}
\begin{center}
\includegraphics[scale=0.75]{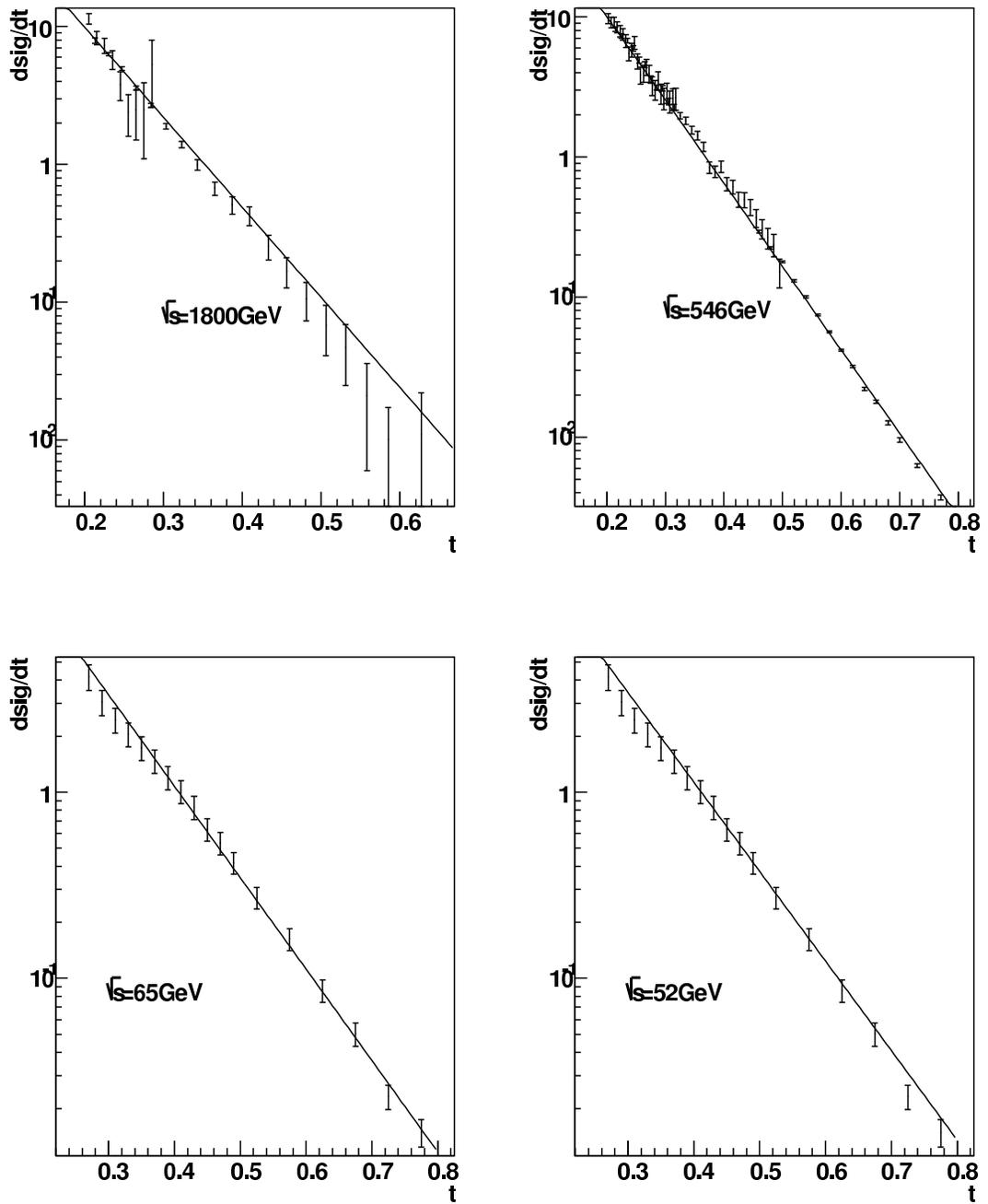}
\end{center}
\caption{$\frac{d\sigma}{dt}$ at $c=0$ (pure pomeron) and at optimal value $c=0.125$. Difference between graphics is not viewable, so we give one figure.}
\label{fig:el_0}
\end{figure}

\begin{figure}
\begin{center}
\includegraphics[scale=0.75]{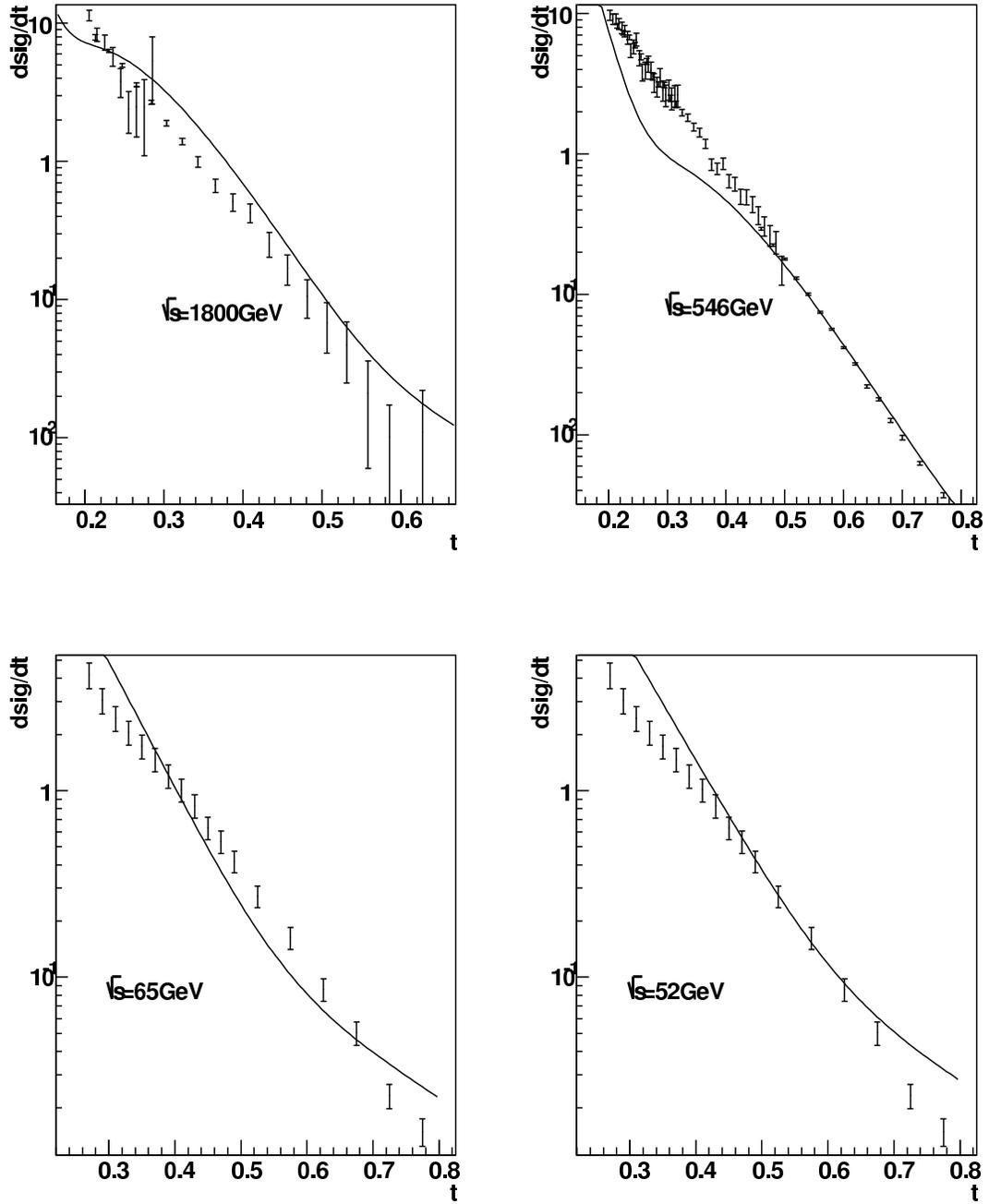}
\end{center}
\caption{$\frac{d\sigma}{dt}$ at $c=0.4$ (region of hardly divergence of experimental data and eikonal model)}
\label{fig:el_04}
\end{figure}

Dependence of $\frac{d\sigma}{dt}$ at some critical values of $c$ is given at Fig.\ref{fig:el_0} and Fig.\ref{fig:el_04}.

\newpage

It is clear, that this eikonal model with exponential pomeron vertex can describe experimental data only at low region $c \leq 0.2$. Including of total cross-section does not change this situation, function $\frac{\chi^2}{n.d.f.}$ rises even more quickly and has no minimum at non-zero $c$. Theory predict, that $c$ is high, $c \sim 1.5$ and there is strong limit $c>1$, so we have serious problem. We can try to reject exponential form of pomeron vertex, but we have no other assumptions about functional form of the vertexes.

It is natural to extract eikonal profile function $\chi(b)$ from experimental data  and test it regge-like behavior To make it, we approximate observed amplitude $A(s,t)$ by regge-like form:
\begin{equation}
\begin{array}{l}
A(s,t)=A(s,0)e^{B(s)t}\\
A(s,0)=s\sigma_{tot}(s)(\rho+i)\\
B(s)=\frac{B_{real}(s)}{2}+i\pi\frac{\alpha^{\prime}}{2}\\
\alpha^{\prime} \sim 0.25
\label{param}
\end{array}
\end{equation}
there $B_{real}(s)$ is logarithmic slope of $\frac{d\sigma}{dt}$ at given $s$. Reliable of this parametrisation is clear from Fig.\ref{fig:el_0}. We have only to test stability of our results on the changes of unobservable imaginary part of $\beta(s)$.

Calculations is trivial enough, and we have for the generic amplitude
\begin{equation}
a(s,\overline{k})=\frac{4\pi s}{ic^2}\int ln \left(1+\frac{ic^2A(s,0)}{8\pi sB(s)}e^{-\frac{b^2}{4B}} \right)J_0(bk)bdb
\end{equation}
Imaginary part of logarithm we define to be small at high $b$ and to be continuous at all $b$.

\begin{figure}
\begin{center}
\includegraphics[scale=0.75]{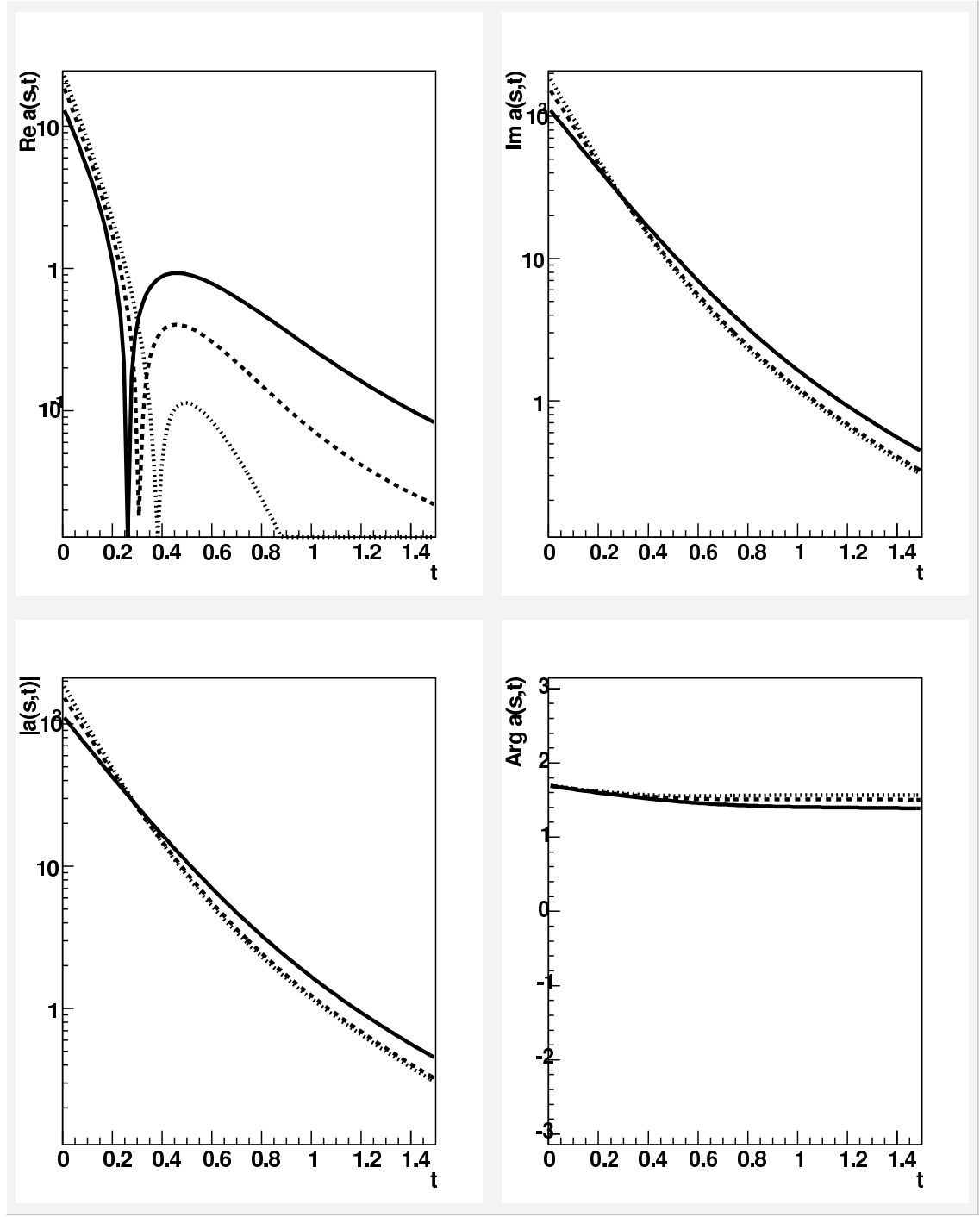}
\end{center}
\caption{Generic amplitude $a(s,t)$, divideded on $s$, reconstructed for c=0.75 at energies  $\sqrt{s}=1800 GeV$ (dotted line),  $\sqrt{s}=546 GeV$ (dashed line), $\sqrt{s}=65 GeV$ (ssolidline)}
\label{gener075}
\end{figure}

\begin{figure}
\begin{center}
\includegraphics[scale=0.75]{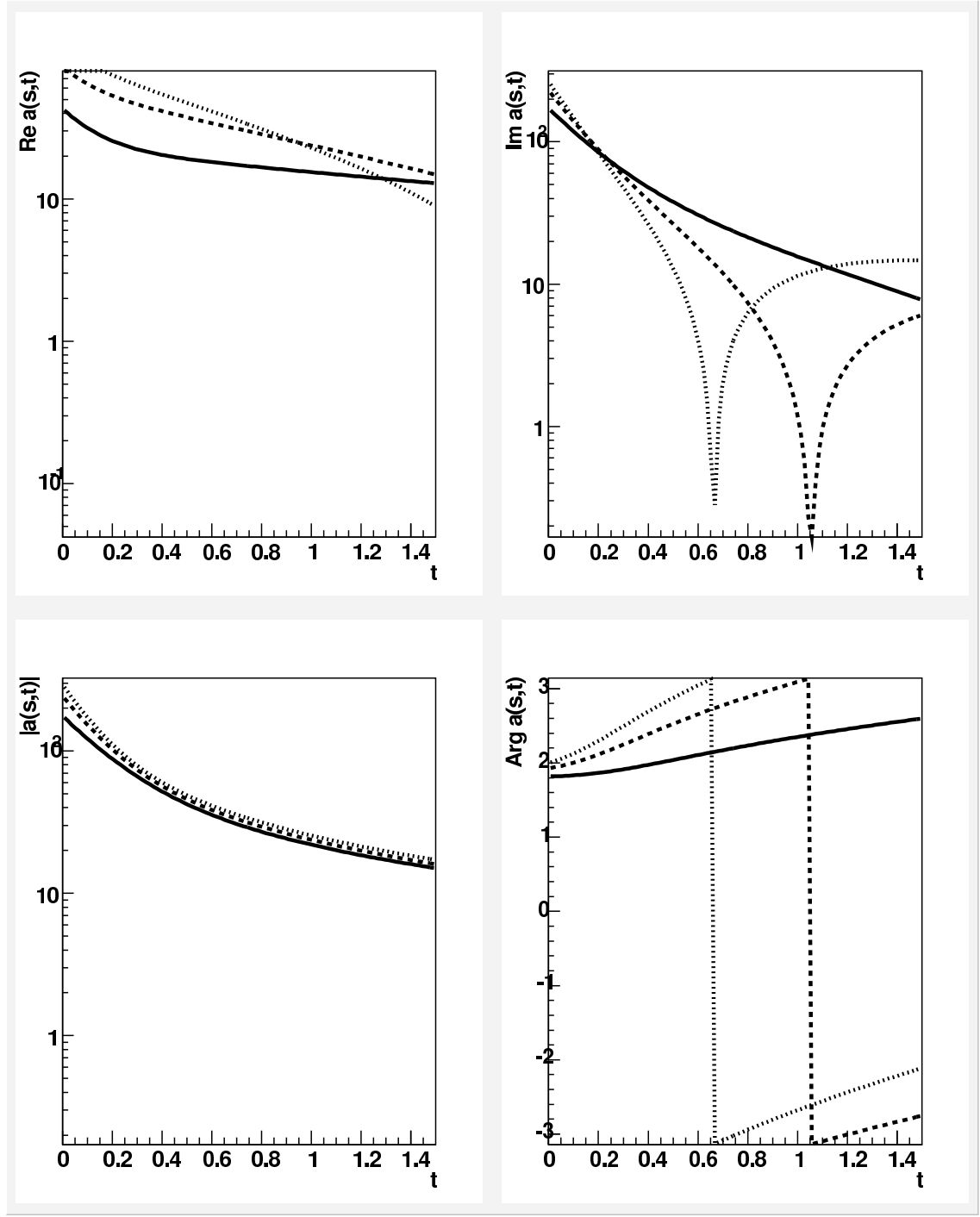}
\end{center}
\caption{Generic amplitude $a(s,t)$, divideded on $s$, reconstructed for c=1.2 at energies  $\sqrt{s}=1800 GeV$ (dotted line),  $\sqrt{s}=546 GeV$ (dashed line), $\sqrt{s}=65 GeV$ (ssolidline)}
\label{gener12}
\end{figure}

Dependence of the reconstructed generic amplitude on $t$ is shown at Figs.\ref{gener075}-\ref{gener12}. We don`t include reconstructed amplitude at trivial $c=0$ here.

\begin{figure}
\begin{center}
\includegraphics[scale=0.7]{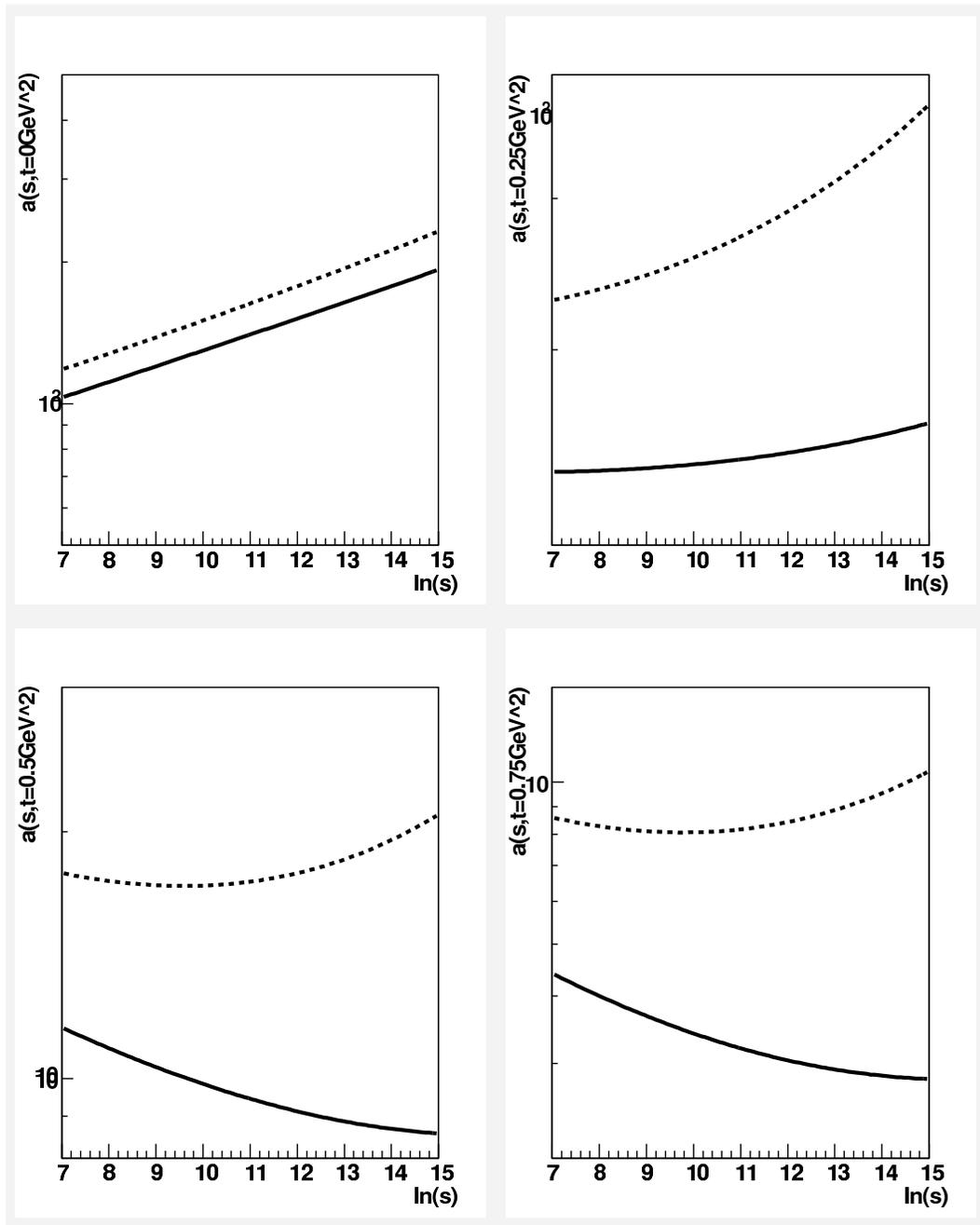}
\end{center}
\caption{Dependence of $a(s,t)$ on $ln(s)$ at $c=0.75$ (solid line) and $c=1$ (dashed line)}
\label{gen_s}
\end{figure}

From the canonical Regge theory, we expect for $a(s,t)$ to be
\begin{equation}
a(s,t)=e^{i\frac{\pi\alpha(t)}{2}}\beta(t)s^{\alpha(t)}
\label{Regge}
\end{equation}
there residue $\beta(t)$ is real.
We see, that our reconstructed amplitudes does not satisfy to equation (\ref{Regge}) at high $c$, because we have $s$-depended phases and (it is more serious), zeroes moving with $s$. Moreover, it is clear from aanalyzeof $s$-dependence of $a(s,t)$ (see Fig.\ref{gen_s}), that $a(s,t)$ does not scale as  $s^{\alpha(t)}$ at high $c$.

\newpage

This results is stable under our assumptions about the phase of $A(s,t)$, at $c>0.75$ we can choose any small imaginary part of $B(s)$ in equations (\ref{param}) without any influence on the reconstructed amplitude $a(s,t)$. It is possible, that $s$-dependence phase of $a(s,t)$ is arisen from wrong assumptions about phase of the $A(s,t)$, but moving zeroes is stable at high $c$ ($c \geq 1$).

We can consider most general properties of the pomeron cuts. Many theoretical models (see \cite{collins} for review) leads us to the next form for the two-pomeron cut
\begin{equation}
\begin{array}{l}
A^c_2(s,t)=\frac{i}{16\pi^2s}\int^{0}_{-\infty}dt_1dt_2\frac{\theta(-\lambda(t,t_1,t_2))}{(-\lambda(t,t_1,t_2))^{1/2}}(N(t,t_1,t_2))^2 A^P(s,t_1)  A^P(s,t_2) \\
\lambda(t,t_1,t_2)=t^2+t_1^2+t_2^2-2(tt_1+tt_2+t_1t_2)
\end{array}
\end{equation}

Amplitude $A^P(s,t)$ is mostly imaginary, so cut $A^c_2$ have opposite sign to pole $A^P(s,t)$. As it was stated above (Figs. \ref{fig:profile_Regge}, \ref{fig:asimpt_I0}), $A^P(s,t)$ is large enough, close to its asimptotic value. So, at high enough $N(t,t_1,t_2)$ cuts will lead to changing of the exponential generic form of the pole (log-plot became non-linear with up-to convexity) and to existence of zeroes, which is moving with $s$. No of this phenomena is observed, so $N(t,t_1,t_2)$ must be small.

Straightforward estimation of eikonal parameter was done by A.Donnachie and P.V.Landshoff at \cite{land_86} and \cite{land_05}, by requiring that the dip discovered in CERN ISR $pp$ elastic scattering occurs at the correct value of $t$. Donnachie-Landshoff value $0.4$ is, actually, the upper bound on eikonal parameter, because we consider soft non-perturbative pomeron eikonalization, but dip at $t \sim 1 GeV^2$ generated by perturbative triple gluon exchange\cite{land_86}, which is out of our work.

So, we can state, that eikonalization parameter $c$ is limited by $c<0.5$, if we assume exponential form of the generic pomeron and by $c<1$ if we make no assumptions about form of the pomeron residue.  Anyway, this results is in contrast with the theory expectations and strongly limit eikonalization models, such as restoring of unitarity and renormalization of the pomeron flux in the single diffraction.

\subsection{Description of the single diffraction data in the framework of the quasi-eikonal model.}
\label{sub:quasi}

Quasi-eikonal model, considered in this work, is standart enough. We use reggeon diagram technic with reggeon propogator $s^{\alpha(t)}$, model gauss vertexes of the interaction of n pomerons with hadron 
\begin{equation}
N_h(k_1,..,k_n)=g_h(g_hc_h)^{n-1}exp\left(-R_h^2\sum_{i=1}^{n}k_i^2\right)
\end{equation}
and the vertex corresponding to the transition of $l$ pomerons into $m$ pomerons under the $\pi$ -meson exchange dominance assumption
\begin{equation}
\Lambda(k_1,..,k_{m})=r(g_{\pi}c_{\pi})^{m-3}exp\left(-R_r^2\sum_{i=1}^{m}k_i^2\right).
\end{equation}
Here $g_h$ is the pomeron-hadron coupling, $c_h$ is the corresponding shower enhancement coefficient, $R_h$ and $R_r$ are
the radii of the pomeron-hadron and pomeron-pomeron interactions, respectively, $k_i$ are the pomeron transverse momenta.
Integration on the pomeron momenta is made trivial in the impact parameter space represenztation and we only have to sum on then nubmers of pomerons, attached to the same vertexes.

\begin{figure}
\begin{center}
\includegraphics[scale=0.5,angle=90]{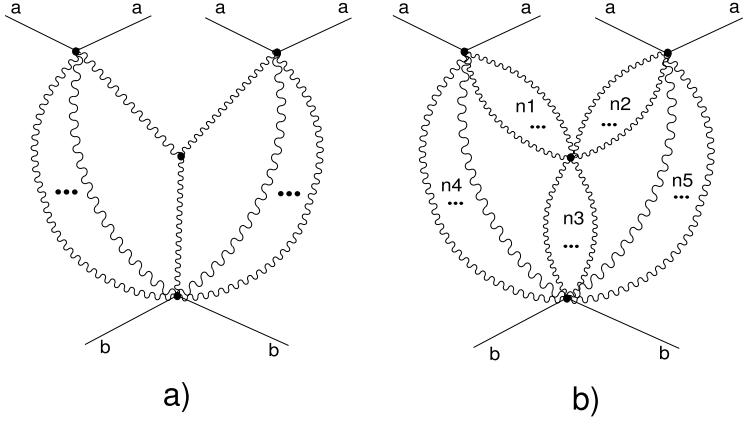}
\vspace*{0pt}
\caption{Regge diagrams describing single diffraction dissociation of particle b.}
\label{fig:fig2}
\end{center}
\end{figure}

As compared with Ref.\cite{gotsman_orig} and Ref.\cite{chung_orig}, where only part of sufficient diagrams  was accounted ( see Fig. \ref{fig:fig2}a), in this paper we account all non-enhanced absorptive corrections to the Y-diagram contribution, shown in  Fig.\ref{fig:fig2}b.

Because low-energy corrections rapidly decrease with energy, we account only pomeron contributions, but in all
sufficient diagrams, as it was done in Ref.\cite{all} and Ref.\cite{Ab}. It gives us possibility to normalize cross-section of the single diffraction to CDF data and make theoreticaly based predictions for cross-section of the single diffraction at LHC energies.

The contribution $f_{n_1 n_2 n_3 n_4 n_5}$ of each diagram in Fig.\ref{fig:fig2}b can be written in a \begin{it} rather simple form \end{it}
 \begin{equation}
\begin{array}{l}
f_{n_1 n_2 n_3 n_4 n_5}=\frac{(-1)^{n_1+n_2+n_3+n_4+n_5+1}}{n_1!n_2!n_3!n_4!n_5!}\frac{8 \pi^3 r}{c_a^2 c_b g_{\pi} c_{\pi}}
\left[\frac{g_a c_a g_{\pi} c_{\pi} e^{\Delta (Y-y)}}
{8\pi (R_a^2+R_{\pi}^2+\alpha^{\prime}(Y-y))}\right]^{n1+n2} \\
\\
\left[\frac{g_a c_a g_b c_b e^{\Delta Y}}
{8\pi (R_a^2+R_b^2+\alpha^{\prime}Y)}\right]^{n4+n5}
\left[\frac{g_b c_b g_{\pi} c_{\pi} e^{\Delta y}}
{8\pi (R_b^2+R_{\pi}^2+\alpha^{\prime}y)}\right]^{n3}
\frac{1}{detF}e^{-t\frac{c}{detF}} \\
\\
detF=a_1 a_2 a_3 + a_1 a_3 a_5 + a_1 a_2 a_5 + a_1 a_2 a_4 + a_2 a_3 a_4 + a_1 a_4 a_5 + a_3 a_4 a_5 + a_2 a_4 a_5 \\
c=a_2 a_3 + a_1 a_5 + a_3 a_5 + a_2 a_5 + a_1 a_3 + a_1 a_4 + a_3 a_4 + a_2 a_4 \\
a_1=\frac{n_1}{R_a^2+R_{\pi}^2+\alpha^{\prime}(Y-y)}\\
a_2=\frac{n_2}{R_a^2+R_{\pi}^2+\alpha^{\prime}(Y-y)}\\
a_3=\frac{n_3}{R_b^2+R_{\pi}^2+\alpha^{\prime}y}\\
a_4=\frac{n_4}{R_a^2+R_b^2+\alpha^{\prime}Y}\\
a_5=\frac{n_5}{R_a^2+R_b^2+\alpha^{\prime}Y}.
\end{array}
\end{equation}
Here $Y=ln(s)$ $y=ln(M^2)$.
Then inclusive cross section is
\begin{equation}
(2\pi)^3 2E \frac{d^3\sigma }{dp^3 }=\pi \frac{s}{M^2}\sum_{n_1,n_2,n_3=1}^{\infty} \sum_{n_4,n_5=0}^{\infty}f_{n_1 n_2 n_3 n_4 n_5}
\end{equation}

Our method differs from early work Ref. \cite{Ab}, where all parameters but vertex $r$ were fixed. In this work here we vary all parameters. Parameters were varied with natural limitations, i.e. all parameters were varied above its conventional values. We don`t consider very high or very low values of pomeron intercept and slope, which can be compensated by other parameters.

Another difference as compared with Ref.\cite{Ab} is the fact, that we use data on total and elestic (differential) cross-sections and data on total single-diffraction cross-sections. So, we can fix parameters of the model with higher precision and with account of its one-to-one corellations.

The model under consideration doesn`t include possible contributions of low-lying reggeons, so we limit considered energies by $\sqrt{s}>52 GeV$ for elastic and total cross-sections. As was shown in \cite{ejela}, modern data don`t give us possibility to distinct simple pole model with total cross-sections  $\sigma_{tot}=As^\Delta$ and eikonaliezed models with  $\sigma_{tot}=C+Dln(s)$ or $\sigma_{tot}=E+Fln(s)^2$. But we can reliably determine parameters of the model $R_h,g_h,\Delta,\alpha^{\prime}$ from elastic and total cross-section data at fixing $c_h$.

We use CDF data on single diffraction for analysis.
 
CDF data \cite{CDF} was presented as a result of the monte-carlo simulations based on the general formula:
\begin{equation}
\begin{array}{l}
\frac{d^2\sigma}{d\xi dt}
=\frac{1}{2}\left[\frac{D}{\xi^{1+\epsilon}}
e^{\textstyle (b_0-2\alpha'_{SD}\ln \xi)t} 
+I\xi^{\textstyle \gamma}e^{\textstyle b't}\right]
\label{CDFfit} \\
\xi \equiv 1-x
\end{array}
\end{equation}

Taken CDF data is shown in Table 1.

\begin{table}[h]
\caption{CDF fit-parameters from reference~[1].}
\begin{center}
\begin{tabular}{|c|c|c|}
\hline
         &                       &                        \\
         & $\sqrt{s}=546\,\,GeV$ & $\sqrt{s}=1800\,\,GeV$ \\
         &                       &                        \\ \hline
         &                       &                        \\
$D$      & $3.53 \pm 0.35$       & $2.54 \pm 0.43$        \\
$b_{0}$  & $7.7 \pm 0.6$          & $4.2 \pm 0.5$          \\
$\alpha'_{SD}$ & $0.25 \pm 0.02$      & $0.25 \pm 0.02$        \\
$\epsilon$ & $0.121 \pm 0.011$   & $0.103 \pm 0.017$      \\
$I$      & $537^{+498}_{-280}$    & $162^{+160}_{-85}$       \\
$\gamma$ & $0.71 \pm 0.22$       & $0.1 \pm 0.16$         \\
$b'$     & $10.2 \pm 1.5$        & $7.3 \pm 1.0$          \\
         &                       &                        \\ \hline
\end{tabular}
\end{center}
\end{table}

This parameters are experimental points tested in our model. Let`s mark, that low-lying reggeons contribution, corresponding to second addendum in  (\ref{CDFfit}), isn`t acconted in our calculations and we have to model only parameters  $D$,$b_{0}$,$\alpha'_{SD}$,$\epsilon$. We calculate this parameters in the region ${0.05<t<0.1; 0.99<x<0.995}$, where we have the most reliable CDF data and contribution of the low-lying reggeons is mnimal.

We have to mark, that this data is not precise because of the following reasons:
 \begin{enumerate}
 \item CDF single diffraction data have low statistcs and narrow kinematical region, where the data was taken;
 \item At each energy ($\sqrt{s}=546GeV$ É $\sqrt{s}=1800GeV$) 6 highly correlated parameters are introduced, and it makes calculations unstable;
 \item Fixing of effective pomeron slope on the common value  $\alpha'_{SD}=0.25$ is obliged.
\end{enumerate}
  
Unreability of the data in Table1 is cearly seen from analysis of dependence of  $D$ on energy from $\sqrt{s}=546GeV$ to $\sqrt{s}=1800GeV$. As defined \cite{CDF}, 
\begin{equation}
D=G(0)s^{\Delta}
\end{equation}
here G(0) doesn`t depend on $s$, and $\Delta>0$. In accordance with this definition, parameter $D$ must increase when energy increases, but in  CDF data it decreases.

Total single diffraction cross-sections are well experimentally defined and don`t depend on the model, used in analysis of basic data (detectors counts)
\begin{equation}
\begin{array}{ll}
\sigma_{SD}(\sqrt{s}=546 GeV)=7.89 \pm 0.33 mb \\
\sigma_{SD}(\sqrt{s}=1800 GeV)=9.46 \pm 0.44 mb
\end{array}
\label{sig_SD_exp}
\end{equation}

We include these two points in  $\chi^2$ test with larger weights, than points shown in Table 1.

Because total and elastic cross sections, on one side, and single diffraction cross sections, on other side, have different types, we vary parameters  $r$,$R_{\pi}$ and $c_{\pi}$ to achieve the best agreement with data in Table1 and data (\ref{sig_SD_exp}), fixing at each step parameters  $\Delta$,$\alpha^{\prime}$,$g_h$ and $R_h$ from total and elastic cross-sections.

Results.

In the end of optimization process we`ve got next parameter set:
$g_p^2=  75.0538,  \Delta=  0.0868089,  R_p^2=1.94755, \alpha\prime=0.148963, 
c_p^2=2.03954,  r=0.111525,  R_{\pi}^2=0.173682,  c_{\pi}^2=6.00989$

Following total single diffraction cross sections were calculated at these parameters:

\begin{equation}
\begin{array}{ll}
\sigma_{SD}(\sqrt{s}=546 GeV)=7.5 mb \\
\sigma_{SD}(\sqrt{s}=1800 GeV)=10 mb
\end{array}
\label{sig_SD_model}
\end{equation}

Corresponding differential characteristics of differential single diffraction cross sections are enumerated in Table 2:

\begin{table}[h]
\caption{Differential characteristics of differential single diffraction cross sections in our model.}
\begin{center}
\begin{tabular}{|c|c|c|}
\hline
         &                       &                        \\
         & $\sqrt{s}=546\,\,GeV$ & $\sqrt{s}=1800\,\,GeV$ \\
         &                       &                        \\ \hline
         &                       &                        \\
$D$      & $2.9628$       & $3.08731$        \\
$b_{0}$  & $5.32553$          & $5.27187$          \\
$\alpha'_{SD}$ & $0.294999$      & $0.270871$        \\
$\epsilon$ & $0.0580572$   & $0.0549202$      \\
         &                       &                        \\ \hline
\end{tabular}
\end{center}
\end{table}

Calculated differential characteristics are very close to ones from the triple-pomerom model, so the following relation is satisfied
\begin{equation}
\frac{d^3\sigma}{dM^2dt}=f_{I\!\!P /p}(x,t)\sigma_{I\!\!Pp}(s)
\end{equation}
where 
\begin{equation}
f_{I\!\!P /p}(x,t)=K(s)\xi^{1-2\alpha_{I\!\!P}(t))}
\end{equation}
is renormalized pomeron flux. As compared with standart triple-pomeron model the dependence of factor $K$ on energy $s$ is introduced. This dependence provides slowing on the rise of the single diffraction cross section with energy.

\begin{figure}
\begin{center}
\includegraphics[scale=0.75]{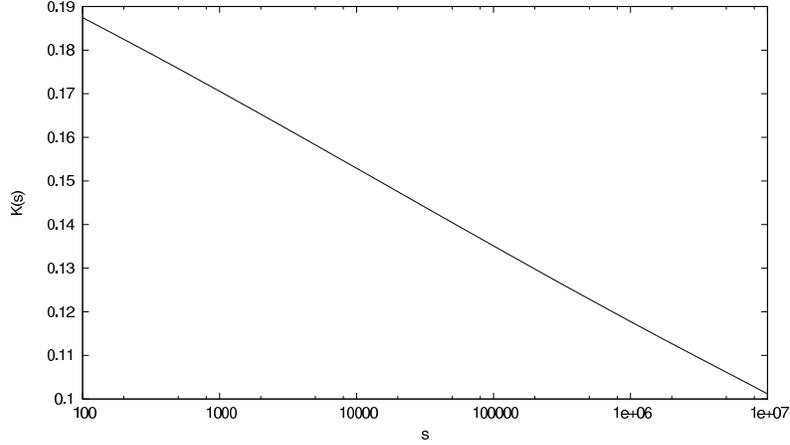}
\vspace*{0pt}
\caption{Dependence of renormalizing factor $K(s)$ on energy.}
\label{fig:K_s}
\end{center}
\end{figure}

Dependence of renormalizing factor $K(s)$ on energy is shown on Fig.\ref{fig:K_s}.

\begin{figure}
\begin{center}
\includegraphics[scale=0.75]{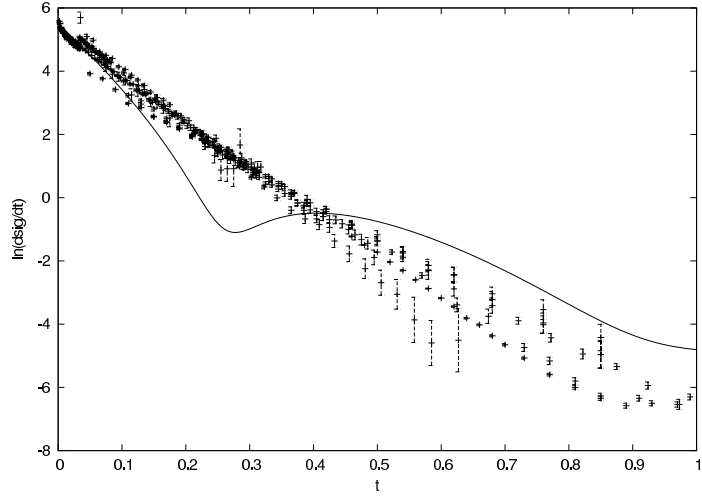}
\vspace*{0pt}
\caption{Elastic cross sections $\frac{d\sigma}{dt}$ for reaction $p+p \rightarrow p+p$. Theoretical curve is at energy $\sqrt{s}=1800GeV$. Experimental points are taken at energies from ISR to Tevatron.}
\label{fig:elastic}
\end{center}
\end{figure}

As we have noted in section \ref{sub:totel} there are inconsistences in esimating $c_p$. From one side, there are theoretical indications, that $c_p>1$. Such high values of $c_p$  lead to significant divergence of dependence $\frac{d\sigma}{dt}$ on $t$ from exponential behavior $e^{-bt}$ already at  $t{\sim}0.2 GeV^2$. It is known from experiment, that elastic cross-section falls exponentially on $t$ up to $t{\sim}1GeV^2$. This inconsistence is clearly seen from Fig.\ref{fig:elastic}.

From this fact of independence of logarithmic slope on $t$ we conclude, that $c_p<<1$.
To explain slow rise of  $\sigma_{SD}$ with energy we have to assume very high $c_{\pi}$, $c_{\pi}c_p \gg 1$ at $g_{\pi} \sim g_{p}$. It gives desired value of the fraction  $\frac{\sigma(\sqrt{s}=1800GeV)}{\sigma(\sqrt{s}=546GeV)} \sim 1.2$, but leads to very high values of logarithmic slope  $b \sim 50 GeV^{-2}$ (situation will be even more worse, than in the case of elastic cross-section, shown on Fig.\ref{fig:elastic}). Solving of this problem by precise adaptation of $R_{\pi}$ is unusable, because it leads to highly differing from experiment and depended on $M^2$ and $t$ values of  $\alpha^{\prime}$ and $\epsilon$.

At $c_p>1$ we don`t need $c_\pi$ in so high values, and logarithmic slope $b$ is back to values about ones, not tens. So, we must return to the theoretically based area  $c_p>1$ and limit considered area of elastic scattering by $t<0.2 GeV^2$.

\begin{figure}
\begin{center}
\includegraphics[scale=0.8]{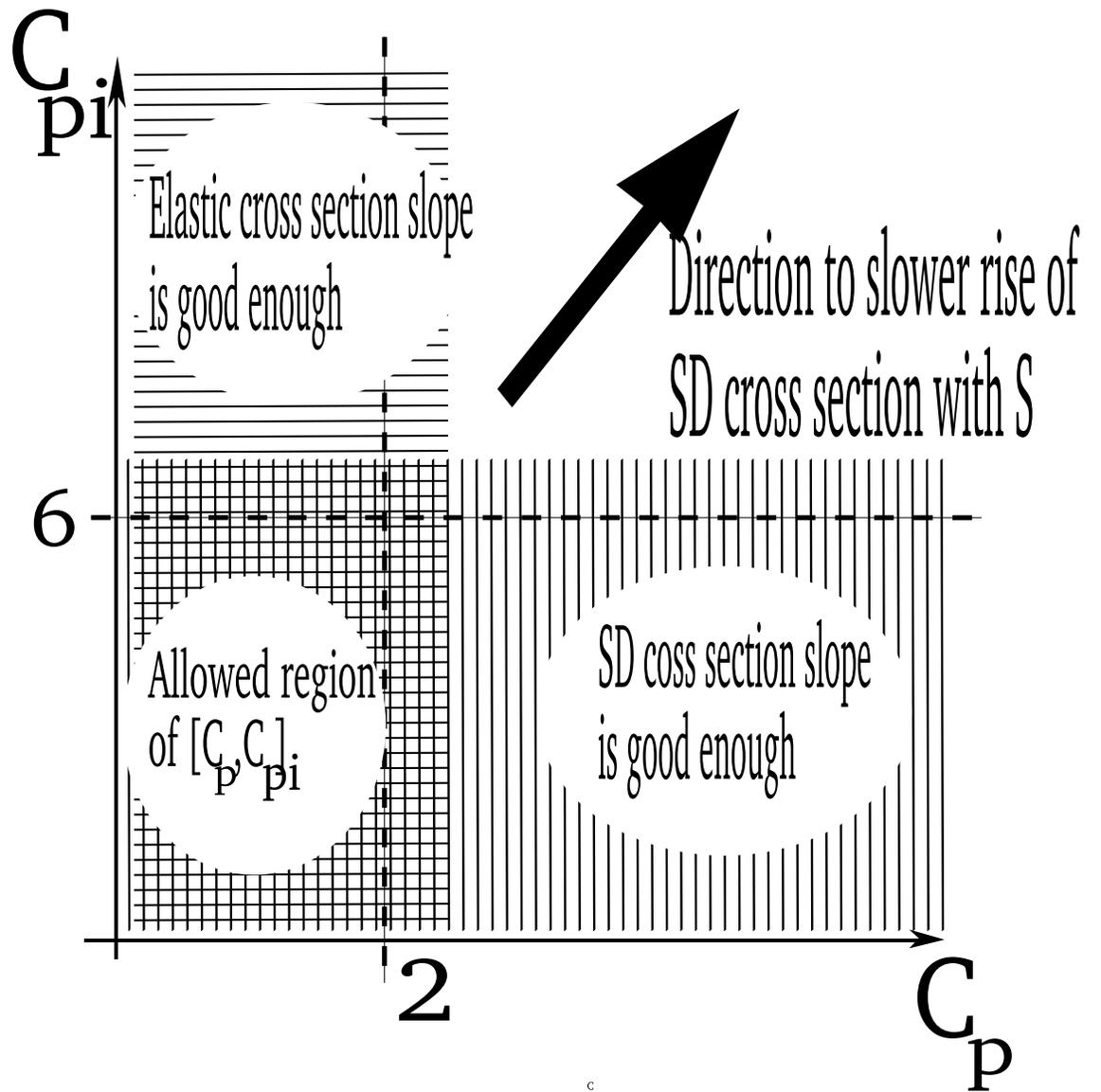}
\vspace*{0pt}
\caption{Allowed region of parameters $[c_p,c_\pi]$.}
\label{fig:call}
\end{center}
\end{figure}

We see, that goog agreement of quasi-eikonal model with experiment is achieved on the border of the allowed region of parameters  $[c_p,c_\pi]$ (see. Fig.\ref{fig:call}).

\newpage

From one side, it gives stability of the calculated parameters. From the other side this model has no reserve of stability. If fraction of the cross sections $\frac{\sigma(\sqrt{s}=1800GeV)}{\sigma(\sqrt{s}=546GeV)}$ will be defined more precisely and will be found in the region $1.1 \div 1.15$ (it is minimal value, which is consistent with existed data), then for description of this data we will be obliged to decline either describing elastic and total cross sections or describing logarithmic slope of the single diffraction on $t$.

The main difficulty in quasi-eikonal description of single diffraction is the same as in elastic scattering, we can not construct good full Pomeron propogator as sum of powers of simple-pole Pomerons. Resulting sum is far from Regge-like behavior, while experiment is very Regge-like with exeption of overall normalisation in single diffraction. 

In the hadron-hadron interaction domain we suggest another description of pomeron flux renormalization.

We start from Regge-based estimation of probabilty (at given impact parameter $\overline{b}$) for pomeron to be emitted with momenta $xP_L,\overline{k}$ and to interact with production of hadron system whith mass $M^2$:
\begin{equation}
P(Regge)=f_{I\!\!P /p}(x,t)P_{I\!\!Pp}(M^2)
\label{SGP1}
\end{equation}

For this value we have good phenomenological triple-pomeron approach with paramters extracted from low-energy fit or from elastic rescattering analisys.

\begin{figure}
\begin{center}
\includegraphics[scale=0.4,angle=270]{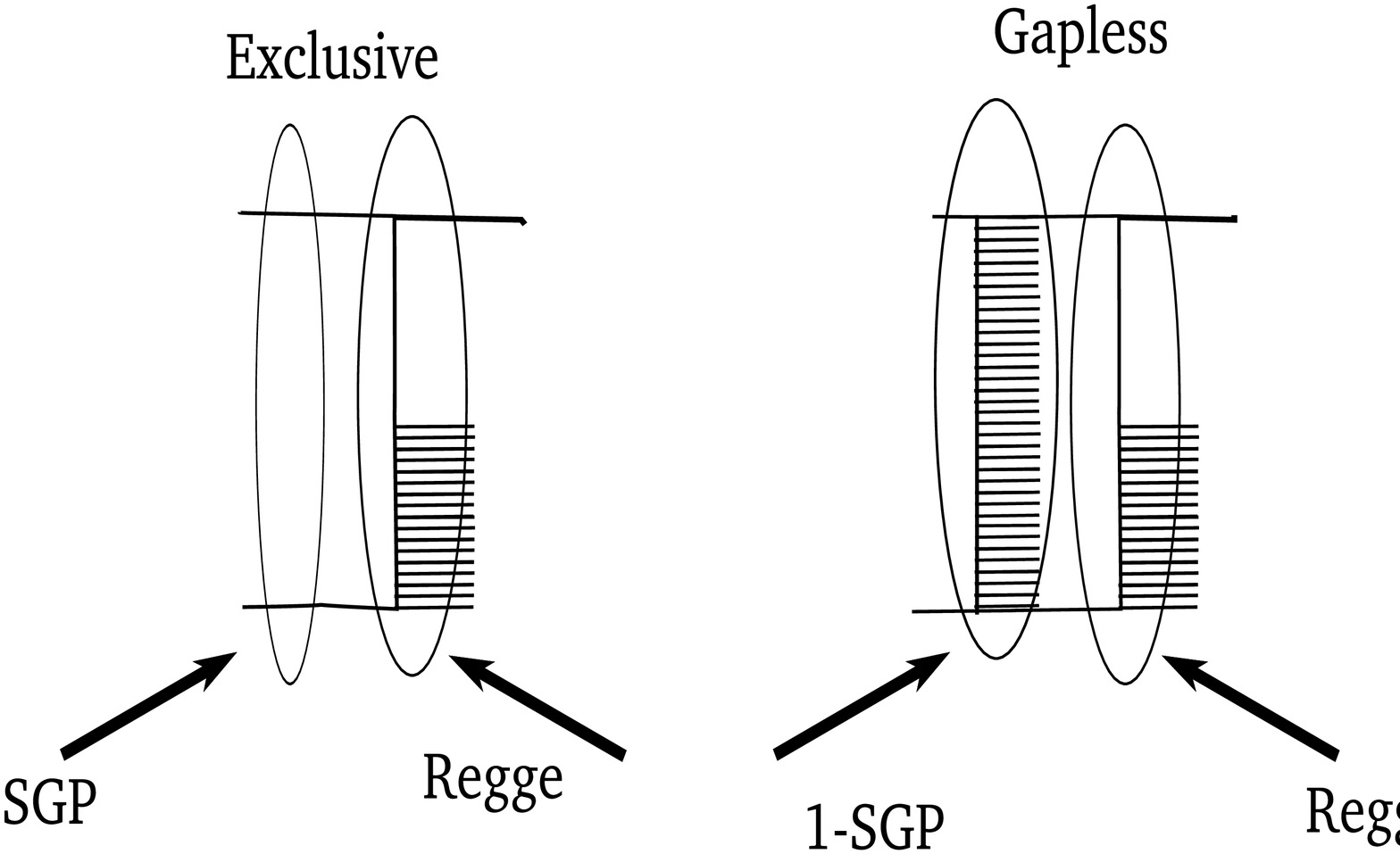}	
\vspace*{0pt}
\caption{SGP corrections to Regge picture of single diffraction.}
\label{fig:SGP_SD}
\end{center}
\end{figure}

Resulting event can include soft rescattering processes, see Fig. \ref{fig:SGP_SD}. 

Surely, we observe only events without soft-rescattering. To estimate rescattering effects, write up (\ref{SGP1}) as sum:
\begin{equation}
P(Regge)=P(Regge|without\_soft\_rescatterings)+P(Regge|with\_soft\_rescatterings)
\label{SGP2}
\end{equation}
Observable is $P(Regge|without\_soft\_rescatterings)$. We assume, that probability of rescattering is independent of Regge part, and, so,
\begin{equation}
P(Regge|without\_soft\_rescatterings)=P(Regge) \times P(without\_soft\_rescatterings)
\label{SGP3}
\end{equation}
Last factor is usually named as survival gap probability (SGP).

Eq.(\ref{SGP3}) leads to estimation
\begin{equation}
\sigma_{SD}(s)=\sigma^{Regge}(s) \times SGP(s)
\label{eq:SD_SGP}
\end{equation}
Term $\sigma^{Regge}(s)$ is triple-pomeron estimation, based on (\ref{SGP1}) or (\ref{triple_gen}).

\begin{figure}
\begin{center}
\includegraphics[scale=0.8]{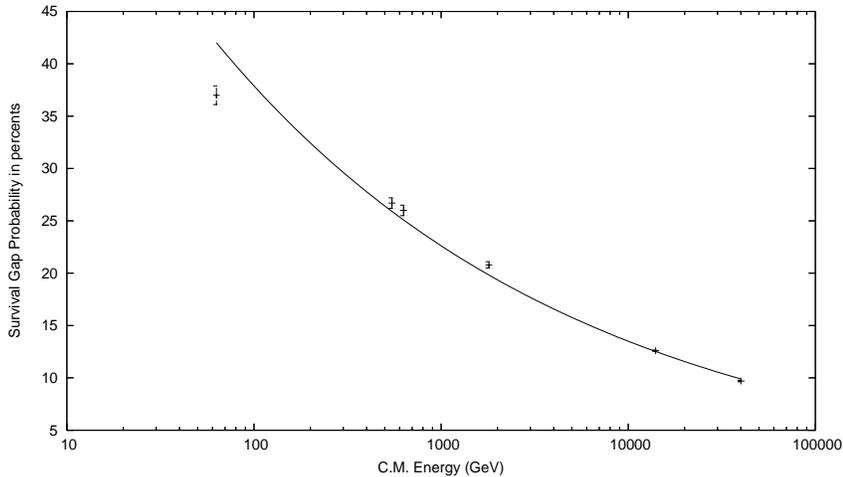}
\vspace*{0pt}
\caption{SGP for higgs production from Ref.\cite{higgs_SGP}.}
\label{fig:SGP_higgs}
\end{center}
\end{figure}

Surely, survival gap probability SGP is much larger for single difraction than for di-jet of higgs production because single diffraction is perefireal. Energy dependence of SGP on $s$ seems to be independent on process, and can be extracted from higgs production estimations. We have used Ref.\cite{higgs_SGP} calculations for dependens SGP on $s$ (see Fig. \ref{fig:SGP_higgs}) and finded
 
\begin{eqnarray}
SGP(s)=Cs^{\epsilon}
\\
\epsilon \sim 0.11
\end{eqnarray}

This value of power is very close to soft Pomeron intercept $\Delta \sim 0.1$. So, it follows from (\ref{eq:SD_SGP}) that observable $\sigma_{SD}(s)$ rise like $ \sim s^\Delta$, not Regge-like $s^{2\Delta}$. 
So, this picture gives good numerical results and can be easily connected to more strict methods, but it fails in description of photon-hadron interactions, where there is no soft rescatterings at all.

\newpage

\section{Applicability of the low constituent model for the description of hadron-hadron diffractive interactions}
\label{sec:lcm}

We consider the three-stage model of hadron interaction at the high energies.

On the first stage before the collision there is a small number of partons in 
hadrons. Their number, basically, coinsides with number of  
valent quarks and slow increases with the rise of energy owing to the appearance 
of the  breasstralung gluons.

On the second stage the hadron interaction is carried out by gluon exchange between the 
valent quarks and initial (bremsstralung) gluons and the hadrons gain the colour 
charge. 

On the third step after the interaction the colour hadrons fly away and when the 
distance between them becomes more than the confinement radius $r_c$, the lines 
of the colour electric field gather into the tube of the radius $r_c$. This tube 
breaks out into the secondary hadrons.

Because the process of the secondary harons production from colour tube goes 
with the probability 1, module square of the inelastic amplitudes corresponds 
to the elastic amplitude with the different number partons.  

\begin{figure}
\includegraphics[width=2.5in]{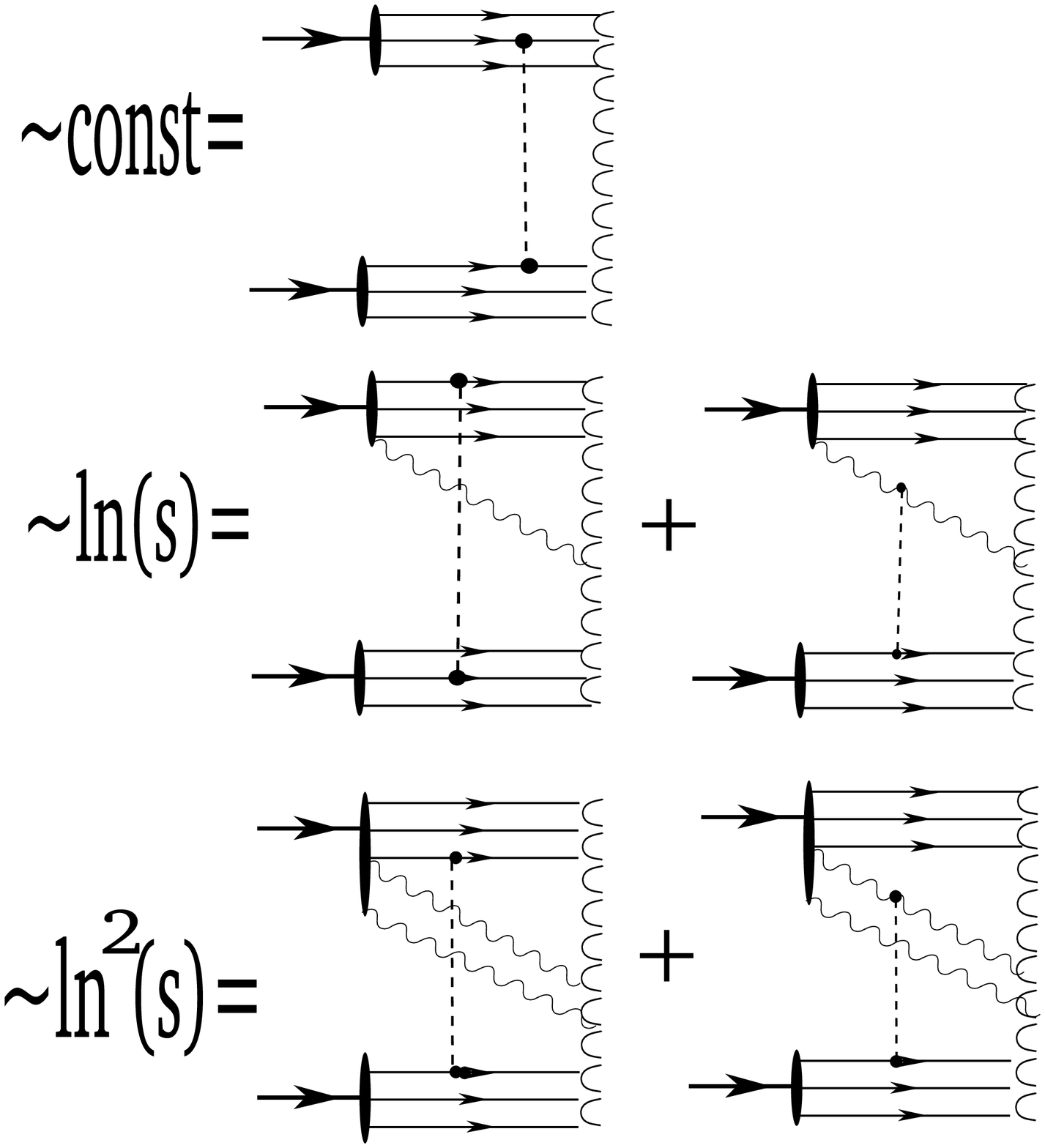}
\vline
\hspace{0.1in}
\includegraphics[width=2.5in]{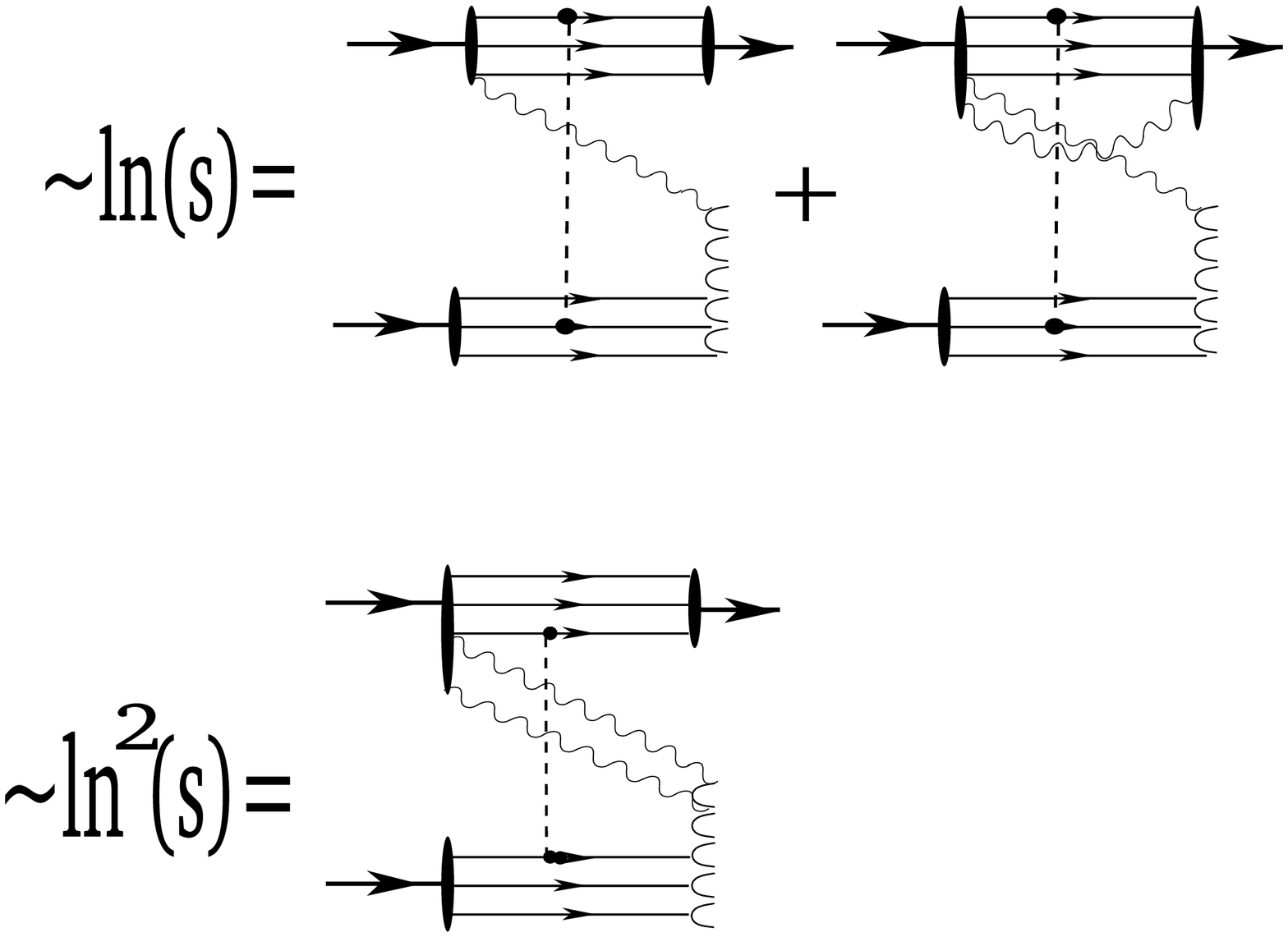}
\caption{Leading by powers of $ln(s)$ diagrams in low constituent model for total cross section (left part) and single diffraction cross section (right part).}
\label{fig:sn}
\end{figure}

This processes are schematically drawn on the left side of Fig.\ref{fig:sn}. 

Processes of the single difraction are naturally discribed in this model. To get colourless final state of the incoming hadron, part of the incoming bremsstralung must be colourless in the sum with exchanged coulomb gluon. So, the part of the longtitudual momenta of the incoming hadron  $x=\frac{M^2}{s}$, which is equal to the sum of $x_i$ exchaged bremsstralung gluons, will be transferred. At the final stage after hadronization we get the system composed of the leading hadron, lost part of the moments $x$, and rapidity separated hadrons with mass $M^2=s(1-x)$. This processes are schematically drawn on the right side of Fig.\ref{fig:sn}.

To estimate the rise of cross section of single diffraction with energy, we will compare leading terms of $ln(s)$ for single diffraction and total cross sectons. Leading diagrams for processes of single diffraction are the same as in the case of the total cross section. To describe processes of single diffraction we must fix colours of exchanged gluons to get colourless final states. This condition cancels constant term in sum of $ln(s)$ powers for single diffraction cross section. In the case of the single diffraction coefficient at $ln(s)$ will be suppressed by factor $\frac{1}{8}$ in comparision to expression for total cross section, because of $t$-channel exchange colourless condition. Coefficient at $ln^2(s)$ will be suppressed by factor $\frac{1}{256}$. Coefficients at different powers of  $ln(s)$ will be also supressed by kinematical factor $K$. It arises because the area of integration on longtitudual momentums of exchanged gluons is limited by the condition, that full exchanged momentum must be in the range $\frac{1.4 GeV^2}{s}..0.15$ by the common definition of $\sigma_{SD}$.  We assume, that these kinematical factors are the same for coefficients at $ln(s)$ and $ln^2(s)$. Therefore, if we take decomposition of cross sections on powers of  $ln(s)$ like
\begin{eqnarray}
\sigma_{tot}=A_1 + B_1 ln(s) + C_1 ln^2(s)\\
\sigma_{SD}=B_2 ln(s) + C_2 ln^2(s)
\end{eqnarray}
then the next conditions take place
\begin{eqnarray}
B_2=\frac{K}{8}B_1 \label{eqn:c1}\\
C_2=\frac{K}{256}C_1 \label{eqn:c2}
\end{eqnarray}
\begin{equation}
F^{color suppresion} \equiv \frac{B_2}{B_1}/\frac{C_2}{C_1}=32
\label{eq:fract}
\end{equation}
Last fraction we can call the colour suppression factor. From the CDF\cite{CDF} experimental data we have $B_1=1.147$,$C_1=0.1466$, and from total cross section data analysis we have $B_2=0.598$, $C_2=0.00214$. For the relation  interested to us (\ref{eq:fract}) we have
\begin{equation}
F^{colour suppresion}_{experimental} \equiv \left( \frac{B_2}{B_1}/\frac{C_2}{C_1} \right) _{experimental} \sim 36
\end{equation}
Precision of the last value is low, but we can state, that $F^{colour suppression}_{min} \sim 10$. Therefore term proportional $ln^2(s)$ is highly suppressed in comparsion with the total cross section, and this effect can be simply explained in our model by existence of color factor in (\ref{eqn:c1})-(\ref{eqn:c2}).

This model gives the same values of pomeron slope for elastic and single diffraction cross sections. This slope value is $\alpha^{\prime} \sim 0.16..0.26$ and consistent with CDF data. This fact is another confirmation of the low constituent model.

\section{From hadron-hadron to photon-hadron diffractive interactions}
\label{sec:photon}

Differences at diffractive properties of hadron-hadron and hadron-photon are traditionally explained in the terms of rescattering effects. Breakdown of factorization from HERA to TeVatron diffractive data is explained by renormalisation of pomeron flux (Goulionas like) either or by low survival gap probability in hadron-hadron interactions at high energies. Anyway, this explanations state that at hadron-hadron interactions rescatterings is high (eikonal parameter $c>1$), and in photon-hadron case rescatterings is suppressed by  factor $1/\alpha_{EM}$.

\begin{figure}
\begin{center}
\includegraphics[scale=0.5]{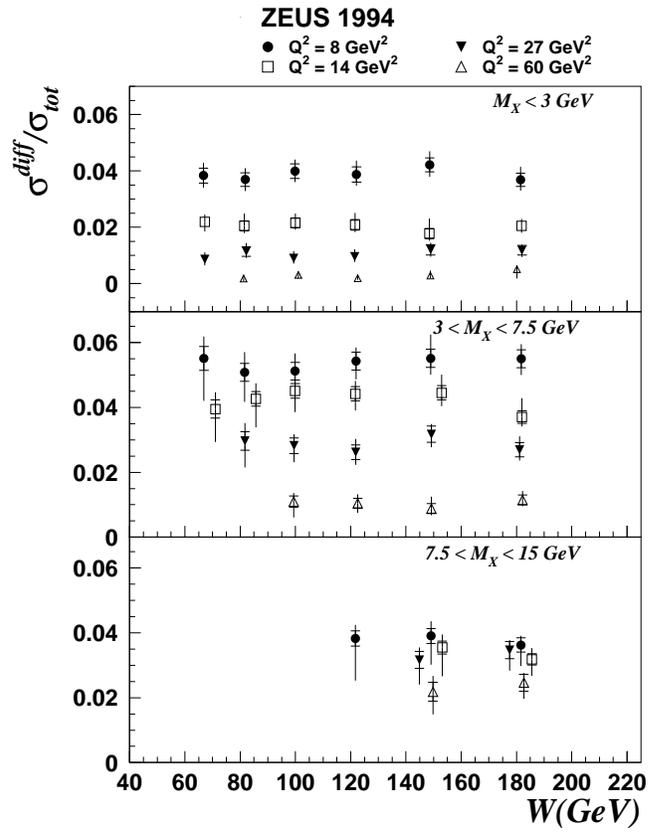}
\label{fig:frac_diff_tot}
\caption{\it Experimantal data for the ratio
$\sigma^{DD}/\sigma_{tot}$ taken from Ref. \cite{ZEUSDATA}.}
\end{center}
\end{figure}

This model gives good description of di-jet diffractive production at TeVatrone compared to HERA one, but failed to explaine constancy of the ratio of diffractive to total cross-sections for photon-hadron interactions at fixed $Q^2$ and $M^2$ (and thus $\beta$ fixed), see Fig.\ref{fig:frac_diff_tot} taken from  \cite{gotsman:rdiff:tot}

Standard triple-Regge theory predicts an increase as $(W^2)^\epsilon$, in clear disagreement with data.

As discussed in \cite{gotsman:rdiff:tot}, no adequate explanation of this fact was found. 

Standart rescattering based solution is failed, because rescattering effects is suppressed.
Saturation-based model (\cite{Golec-Biernat:1999}, \cite{Buchmuller} and references therein) is half-phenomenological and saturation does not seem to be adequate in this wide region of masses $W^2$ and $M_X^2$.

In our low constituent model there is no difference in describing hadron-hadron and photon-hadron single diffraction, because color suppression factor in equation (\ref{eq:fract}) does not depend on the nature of diffracting systems.

\newpage

\section{Acknowledgementes}
This work was supported by grants RFBR 03-02-16157-a, RFBR 05-02-08266-ofi\_a and grant of Ministry of Education 8153.


\begin{thebibliography}{100}

\bibitem{gotsman:rdiff:tot}
{Gotsman E et al, {\em Energy dependence of $\sigma^{DD}/\sigma_{tot}$ in
  DIS\/}, [{\tt hep-ph/0007261}]}
  
\bibitem{ZEUSDATA}
ZEUS collaboration: J. Breitweg et al., {\it Eur. Phys. J.} {\bf C6}
(1999) 43. 

\bibitem{Golec-Biernat:1999}
{Golec-Biernat K and Wusthoff M, PRD {\bf59} (1999) 014017, {\bf 60} (1999) 114023}

\bibitem{Buchmuller}
Buchm\"uller W, {\em Towards the theory of
  diffractive DIS\/}, talk presented at ``New Trends in HERA Physics'',
  Ringberg Workshop, 1999, [{\tt hep-ph/9906546}]
  

\bibitem{higgs_SGP}
  M.~M.~Block and F.~Halzen,
  Phys.\ Rev.\ D {\bf 63} (2001) 114004
  [arXiv:hep-ph/0101022].


\bibitem{fruassar}
  M.~Froissart,
  Phys.\ Rev.\  {\bf 123} (1961) 1053.

\bibitem{Pancheri:2004ct}
G.~Pancheri, Y.~Srivastava and N.~Staffolani,
arXiv:hep-ph/0406321.

\bibitem{land_86}
A Donnachie and P V Landshoff, Nuclear Physics B267 (1986) 690


\bibitem{land_05}
  P.~V.~Landshoff,
  arXiv:hep-ph/0509240.

\bibitem{goul_orig} K. Goulianos, Phys. Lett. B 358 (1995) 379; Erratum: ib. B363 (1995) 268

\bibitem{gotsman_orig}
E. Gotsman, E.M. Levin and U.Maor, Phys. Rev. D 49 (1994) R4321

\bibitem{chung_orig}
Chung-I Tan Phys.Rept.315:175-198,1999

\bibitem{DL} A.~Donnachie and P.~V.~Landshoff,
Nucl.\ Phys.\ B {\bf 231} (1984) 189.

\bibitem{elfig}
H.~G.~Dosch, C.~Ewerz and V.~Schatz,
Eur.\ Phys.\ J.\ C {\bf 24}, 561 (2002)
[arXiv:hep-ph/0201294].

\bibitem{goul_pic}
K.~Goulianos and J.~Montanha,
Phys.\ Rev.\ D {\bf 59}, 114017 (1999)
[arXiv:hep-ph/9805496].

\bibitem{erhan_orig} P. Schlein, Proceedings of the 3rd Workshop on Small-x and Diffractive Physics, Argonne National Laboratory, USA, 26-29 September 1996
\\
S.~Erhan and P.~E.~Schlein,
Phys.\ Lett.\ B {\bf 427}, 389 (1998)
[arXiv:hep-ph/9804257].

\bibitem{collins}
P.~D.~B.~Collins, An Introduction to Regge Theory and High Energy Physics, Cambridge University  Press (1977).

\bibitem{DLfig}
P.~V.~Landshoff,
arXiv:hep-ph/9605383.

\bibitem{all} V.A.Abramovsky Pis`ma v JETF, 23 (1976) 228.
A.Capella, J.Kaplan, J.Tran Thanh Van, Nucl. Phys. B105 (1976) 333.

\bibitem{Ab} V.A.Abramovsky, R.G.Betman Proc.XXIVth Recontre de Moriond, 1989, p.91.

\bibitem{ejela}
J.~R.~Cudell, V.~Ezhela, K.~Kang, S.~Lugovsky and N.~Tkachenko,
Phys.\ Rev.\ D {\bf 61}, 034019 (2000)
[Erratum-ibid.\ D {\bf 63}, 059901 (2001)]
[arXiv:hep-ph/9908218].

\bibitem{CDF} F. Abe {\em et al.}, CDF Collaboration,
Phys. Rev. {\bf D 50} (1994) 5535.

\end{thebibliography}
\end{document}